\documentclass[]{aastex631}

\accepted{April 11, 2022}
\submitjournal{ApJ}

\shorttitle{Fastest Rotating O-type Stars}
\shortauthors{Shepard et al.}

\begin{document}

\title{Spectroscopic Line Modeling of the Fastest Rotating O-type Stars}

\correspondingauthor{Katherine Shepard}
\email{shepard@chara.gsu.edu}

\author[0000-0003-2075-5227]{Katherine Shepard}
\affiliation{Center for High Angular Resolution Astronomy and 
Department of Physics and Astronomy, Georgia State University, 
P.O. Box 5060, Atlanta, GA 30302-5060, USA} 

\author[0000-0001-8537-3583]{Douglas R. Gies}
\affiliation{Center for High Angular Resolution Astronomy and 
Department of Physics and Astronomy, Georgia State University, 
P.O. Box 5060, Atlanta, GA 30302-5060, USA} 

\author[0000-0001-8025-8981]{Lex Kaper}
\affiliation{Anton Pannekoek Institute for Astronomy, 
University of Amsterdam, Science Park 904, Postbus 94249, 1098 GE Amsterdam, \\
The Netherlands}

\author[0000-0002-1198-3167]{Alex De Koter}
\affiliation{Anton Pannekoek Institute for Astronomy, 
University of Amsterdam, Science Park 904, Postbus 94249, 1098 GE Amsterdam, \\
The Netherlands}


\begin{abstract}

We present a spectroscopic analysis of the most rapidly rotating stars currently known, VFTS~102 ($v_{e} \sin i = 649 \pm 52$ km~s$^{-1}$; O9: Vnnne+) and VFTS~285 ($v_{e} \sin i = 610 \pm 41$ km~s$^{-1}$; O7.5: Vnnn), both members of the 30~Dor complex in the Large Magellanic Cloud. 
This study is based on high resolution ultraviolet spectra from HST/COS and optical spectra from VLT X-shooter plus archival VLT GIRAFFE spectra.
We utilize numerical simulations of their photospheres, rotationally distorted shape, and gravity darkening to calculate model spectral line profiles and predicted monochromatic absolute fluxes. 
We use a guided grid search to investigate parameters that yield best fits for the observed features and fluxes. 
These fits produce estimates of the physical parameters for these stars (plus a Galactic counterpart, $\zeta$~Oph) including the equatorial rotational velocity, inclination, radius, mass, gravity, temperature, and reddening. 
We find that both stars appear to be radial velocity constant. 
VFTS~102 is rotating at critical velocity, has a modest He enrichment, and appears to share the motion of the nearby OB association LH 99.  These properties suggest that the star was spun up through a close binary merger. 
VFTS~285 is rotating at $95\%$ of critical velocity, has a strong He enrichment, and is moving away from the R136 cluster at the center of 30~Dor.  It is mostly likely a runaway star ejected by a supernova explosion that released the components of the natal binary system. 

\end{abstract}

\keywords{Stellar rotation (1629); Massive stars (732); Large Magellanic Cloud (903)}

\section{Introduction} 

We now understand that the lives of massive stars depend critically on both stellar mass and rotation, and the evolutionary paths and surface abundances of rapidly rotating stars are radically different from those of slow rotators.
\citet{Ekstrom2008},
\citet{Brott2011},
\citet{Georgy2013},
\citet{Groh2019},
\citet{Murphy2021},
\citet{Eggenberger2021}, 
and others have presented grids of evolutionary tracks for massive stars
of varying mass, rotation rate, and abundance.  They 
generally find that  massive rapidly rotating stars (equatorial velocities greater than 
$\approx 500$ km~s$^{-1}$) become brighter and hotter through their H-core burning lifetime, rather than the usual stellar cooling
associated with evolution towards the red supergiant branch. 
This behavior is due to the extreme 
rotationally-induced mixing that occurs in the interiors of rapidly rotating stars
which transports hydrogen fuel into the core and brings 
processed helium towards the surface. 
The result of this homogeneous evolution through mixing is that the star will continue to
move up the main sequence until the entirety of the internal hydrogen
supply is depleted. Evidence for this form of evolution is an observed enhancement of the helium and nitrogen abundances,
which are indicators of the CNO nuclear burning process actively
occurring in the core of the star \citep{Roy2020}.

How massive stars attain such fast rotation rates is a subject of considerable debate.  
Stars might be born with an inherently large angular momentum while 
others may experience a spin up through interactions in 
close binary systems, a common occurrence among the massive star population.
\citet{deMink2013,deMink2014} argue that many main sequence stars were spun up through processes that transform binary orbital angular momentum into the spin angular momentum of the components.
Very close binary systems may begin interacting during the components' core
H-burning stage, and depending on the circumstances,
may ultimately merge during a common envelope event.  
The merger product may appear as a rejuvenated, rapidly 
spinning single star.  Binary systems with larger separations 
may instead interact at a later evolutionary stage in which steady mass transfer 
can lead to the stripping of the mass donor and the spin up
of the mass gainer \citep{Wellstein2001}.  

Investigating the origins of massive rapid rotators 
requires careful analysis of their spectra including accounting
for the physical changes in stellar properties with rotation. 
The primary measurement from the Doppler broadening of the 
spectral lines is the projected equatorial velocity, $v_e \sin i$.
The inclination $i$ can be directly measured for nearby stars
through long baseline interferometry \citep{Che2011}, but 
otherwise we must rely on subtle changes in the predicted 
spectral line shapes with inclination in order to determine
the equatorial velocity $v_e$ from $v_e \sin i$. 
This requires the use of a spectrum synthesis code that 
performs a numerical integration of the predicted flux emanating from the
visible hemisphere of a rotationally distorted star.  
If we divide the surface of the star into a grid, each surface element contributes a spectral flux increment 
that is the product of its projected area and the Doppler-shifted
specific intensity $I_\lambda$ which is a function of the local effective temperature, surface gravity, atmospheric abundance, 
and the cosine of the angle between the line of sight and 
the surface normal.  The summation of all the flux spectra 
increments yields a model line profile that can be directly 
compared with observations. 

At high rotation rates, the equatorial radius of the star grows
and the polar radius decreases.  The result is a systematic 
temperature variation from the hotter pole to the cooler equator 
that is known as gravity darkening.  The apparent brightness
of a star experiencing gravity darkening will depend on the
orientation of the star relative to the observer's line of sight. 
If the star is oriented closer to pole-on, $i=0^{\circ}$, then
the star will be brighter overall and the measured $v_e \sin i$ will be small. 
If the star is oriented more equator-on, $i=90^{\circ}$,
the overall brightness will be darker and the measured 
projected rotational velocity larger.  However, because the 
equatorial zone contributes to the largest Doppler shifts 
but with relatively less flux, the true rotational velocity 
can be underestimated unless the gravity darkening is 
modeled accurately \citep{Townsend2004}.  The traditional approach
relies on the von Zeipel law \citep{vonZeipel1924} in which the
local temperature varies with colatitude $\theta$ as a power law 
of the local effective gravity,  $T_{{\rm eff}}(\theta) \propto
g_{{\rm eff}}^{\beta}(\theta)$, where $\beta=0.25$ for stars with
radiative envelopes. More recent work by \citet{EspinosaLara2011,EspinosaLara2013}
demonstrates the importance of dealing with the interior structure
of rotating stars in defining the surface temperature variation. 
They present an $\omega$-model as an analytical approximation 
of the results from detailed numerical models. This 
$\omega$-model predicts a smaller difference between the
polar and equatorial temperatures than does the von Zeipel law.

A spectrum synthesis analysis based upon the von Zeipel law 
for gravity darkening was made by \citet{Howarth2001},
who investigated three stars within our galaxy that, at
the time, were the most rapidly rotating stars known 
with $v_e \approx 430$ km~s$^{-1}$ and 
$\Omega / \Omega_{c} \approx 0.9$: 
HD~93521 (O9.5:~V), HD~149757 ($\zeta$ Oph, O9.5:~V), and 
HD~191423 (ON9:~III~n). 
Here $\Omega$ is the angular velocity at the stellar equator
and $\Omega_c$ is the critical angular velocity, or the 
Keplerian angular velocity in the Roche model (with the equatorial
radius equal to $1.5\times$ the polar radius; \citealt{Rieutord2016}).
They utilized a grid of hydrostatic, plane-parallel,
H and He, non-LTE model atmospheres generated by the code TLUSTY 
to create specific intensity spectra for flux integration. 
\citet{Howarth2001} found that all three stars have an 
atmospheric He abundance that is about twice the solar value, 
providing strong evidence of rotationally-induced internal mixing. More recently, the record holder for the fastest rotating 
massive star in the Galaxy was passed to the star 
LAMOST J040643.69+542347.8 (O6.5:~Vnnn(f)p).  
\citet{Li2020} discovered that this is a runaway star with 
a projected rotational velocity of $v_e \sin i = 540$ km~s$^{-1}$.
This will be a key object for future high resolution 
spectroscopy and spectrum synthesis analysis to 
determine its true equatorial velocity.

The fastest rotating stars known today were discovered in 
the massive star forming region of 30~Doradus in the 
Large Magellanic Cloud (LMC).  The VLT-FLAMES Tarantula Survey
(VFTS; \citealt{Evans2011}) is a large spectroscopic 
survey of over 800 massive stars in this region that 
has led to numerous investigations of stellar properties. 
\citet{Ramirez2013} published a plot showing the distribution of
projected rotational velocities among O-stars in the 30 Dor region
(see their Fig.~11). Their histogram of $v_e \sin i$ shows a 
general decline with increasing rotational velocity that reaches zero near $v_e \sin i = 520$ km~s$^{-1}$. However, in several 
velocity bins beyond this, they find two extremely rapid 
rotators, the stars VFTS~102 and VFTS~285, with estimated 
projected velocities of 610 and 609 km~s$^{-1}$, respectively. 
These two stars are the subject of this paper. 

\citet{Dufton2011} were the first to point out the 
extraordinary nature of VFTS~102 (O9:~Vnnne+; 
\citealt{Walborn2014}) and 
to discuss its possible origin.  They measured the 
widths and radial velocities of its very broadened 
and shallow He absorption lines (see their Fig.~1) 
and estimated its physical properties.   
Their derived radial velocity is lower than that
of other neighboring massive stars, and as a result they suggested
that VFTS~102 is a runaway star.  Furthermore, they 
showed that a nearby pulsar PSR~J0537-6910 displays 
an X-ray emitting bow-shock that points back to 
the general direction of VFTS~102. This led 
them to suggest that VFTS~102 is the survivor of a 
supernova explosion in a binary system that led to the ejection of the pulsar. 

The spectrum of the second rapid rotator, VFTS~285
(O7.5:~Vnnn), was first described by \citet{Walborn2012,Walborn2014}.
\citet{Walborn2014} show (in their Fig.~6) the blue portion of the
spectrum of VFTS~285 in relation to other rapidly rotating 
O-stars from the VFTS survey.  The He lines of VFTS~285 are so astoundingly
wide and shallow compared to those of the other spectra, that the 
authors appended an exclamation mark to its labelled 
spectral classification to highlight its remarkable nature. 
The star's basic properties were estimated by 
\citet{Sabin2017}, and its astrometric motion suggests 
that it is a runaway from the central R136 cluster 
\citep{Platais2018}. 

\citet{Shepard2020} described the first ultraviolet spectra 
of VFTS~102 and VFTS~285 that were obtained with the 
Cosmic Origins Spectrograph on the Hubble Space Telescope 
(discussed further in this work). 
They found that the hotter star, VFTS~285, has a two component 
stellar wind. 
The \ion{N}{5} $\lambda\lambda 1238, 1242$ doublet 
shows a fast, sparse outflow associated with the hotter polar
regions, while the \ion{Si}{4} $\lambda\lambda 1393, 1402$
lines show a slower, but dense outflow
associated with the cooler, equatorial zone. 
They found no P~Cygni wind features in the UV spectrum of VFTS~102,
but they confirmed the existence of a circumstellar disk 
which is indicated by the double-peaked emission of the 
H Balmer lines (especially H$\alpha$) and the Paschen series.

Here we investigate the UV and optical spectra (\S2) of VFTS~102 
and VFTS~285 to determine their rotational properties 
and other physical parameters.  We first present 
new radial velocity measurements (\S3) that indicate 
that both stars are radial velocity constant and 
probably single.  We then describe our spectrum synthesis 
code (\S4) that we use to calculate model spectra for 
sixteen lines of interest.  These models are compared to 
observed spectral profiles (\S5) in order to derive the
rotational velocities and other parameters.  We compare 
our results to models of single star and binary star 
evolution (\S6) to explore the possible origins of
these extreme stars.  Our conclusions are summarized in \S7.

\section{Observations} 

Our sample of observations consists of both 
far-ultraviolet (FUV) and optical spectra for VFTS~102, 
VFTS~285, and a Galactic counterpart, $\zeta$~Oph. 
We include an analysis of the spectra of $\zeta$~Oph 
as a check on our methods in comparison to the 
corresponding work by \citet{Howarth2001} and as
Galactic comparison benchmark for considering the 
results for the two LMC stars. 

\subsection{FUV}

We obtained high resolution spectra of VFTS~102 and VFTS~285 
with the Cosmic Origins Spectrograph (COS) on board the 
Hubble Space Telescope (HST).  Comparable spectra of $\zeta$~Oph
were collected from the archive of the International Ultraviolet
Explorer (IUE).

HST/COS is a high dispersion spectrograph designed to record the
FUV spectra of faint point sources \citep{Green2012, Fischer2019}.
The observations reported here were obtained during Cycle~23 as a
part of the program GO-14246. The observations of VFTS~102 were
made over a series of three orbits on 2017 January 1, while the
observations of VFTS~285 were obtained during one orbit on 2016
April 10. These FUV spectra were all obtained using the G130M
grating in order to record the spectrum over the range from 1150
to 1450 \AA\ with a spectral resolving power of
$R=\lambda/\triangle\lambda=18000$. The two detectors on COS are
separated by a small gap, therefore the central wavelength was
varied slightly between observations (1300, 1309, and
1318~\AA ) for VFTS~102 in order to fill in the missing flux. In
each of these settings, four sub-exposures were obtained at four
{\tt FP-POS}, or focal plane offset positions, in order to avoid
fixed-pattern problems. The VFTS~285 spectra were made using the
same method except only two central wavelength positions were
selected, 1300 and 1318~\AA , due to orbital time restrictions. 
The spectrograph parameters are summarized in Table~1. 

\placetable{tab1}      
\begin{deluxetable*}{ccccccccc}
\tabletypesize{\scriptsize}
\movetableright=0.1mm
\tablenum{1}
\tablecaption{Overview of Spectroscopic Observations\label{tab1}}
\tablewidth{0pt}
\tablehead{
\colhead{Spectrograph} &
\colhead{Mode} &
\colhead{Wavelength Range} &
\colhead{Resolving} &
\colhead{$S/N$} &
\colhead{$n$} & 
\colhead{$n$} & 
\colhead{$n$} & 
\colhead{PI} \\
\colhead{} &
\colhead{} &
\colhead{(\AA\ )} &
\colhead{Power} &
\colhead{(pixel$^{-1}$)} &
\colhead{VFTS~102} & 
\colhead{VFTS~285} & 
\colhead{$\zeta$ Oph} & 
\colhead{Name}
}
\startdata
HST/COS       & G130M       & 1150-1440  &18000 & 5  & 1 & 1    & 0  & Gies \\
IUE           & SWP/High    & 1150-2000  &10000 & 4  & 0 & 0    & 72 & Bolton $+$      \\
X-shooter-UVB & Slit 0.5x11 & 3350-5500  & 9700 & 48 & 4 & 7    & 0  & Gies, Przybilla \\
X-shooter-VIS & Slit 0.9x11	& 5540-10150 & 8900 & 70 & 2 & 7    & 0  & Gies, Przybilla \\
X-shooter-VIS & Slit 0.4x11	& 5540-10150 &18400 & 70 & 2 & 0    & 0  & Gies, Przybilla \\
X-shooter-NIR & Slit 0.6x11	& 9959-20800 & 8100 & 47 & 5 & 7    & 0  & Gies, Przybilla \\
GIRAFFE-UV 	  & Medusa/LR	& 3950-4570	 & 6300	& 50 &32 & 34   & 0	 & Evans \\
GIRAFFE-VIS   & Medusa/LR	& 4500-5070	 & 7500	& 35 & 3 & 4    & 0	 & Evans \\
GIRAFFE-NIR   & Medusa/LR	& 6439-6820	 &17000 & 44 & 2 & 4    & 0	 & Evans \\
ESPaDOnS      & SpecPolar   &3700-107500 &68000 & 9  & 0 & 0    &146 & Wade \\
\enddata
\end{deluxetable*}

The HST/COS observations were processed using the standard COS pipeline, merged
onto a single barycentric wavelength grid, and transformed onto a
uniform wavelength grid. The resulting spectra have a signal-to-noise
ratio of $S/N = 5$ per pixel in the central, best exposed regions. For
additional information on this procedure see \citet{Shepard2020}. The FUV
spectra for all three stars are illustrated in Figure~1.  The primary components are
strong Ly$\alpha$ absorption (interstellar), 
the \ion{N}{5} and \ion{Si}{4} wind lines, 
numerous sharp interstellar lines, and shallow blends 
of photospheric lines. 

\begin{figure*}
\centering
\includegraphics[angle=90,width=15cm]{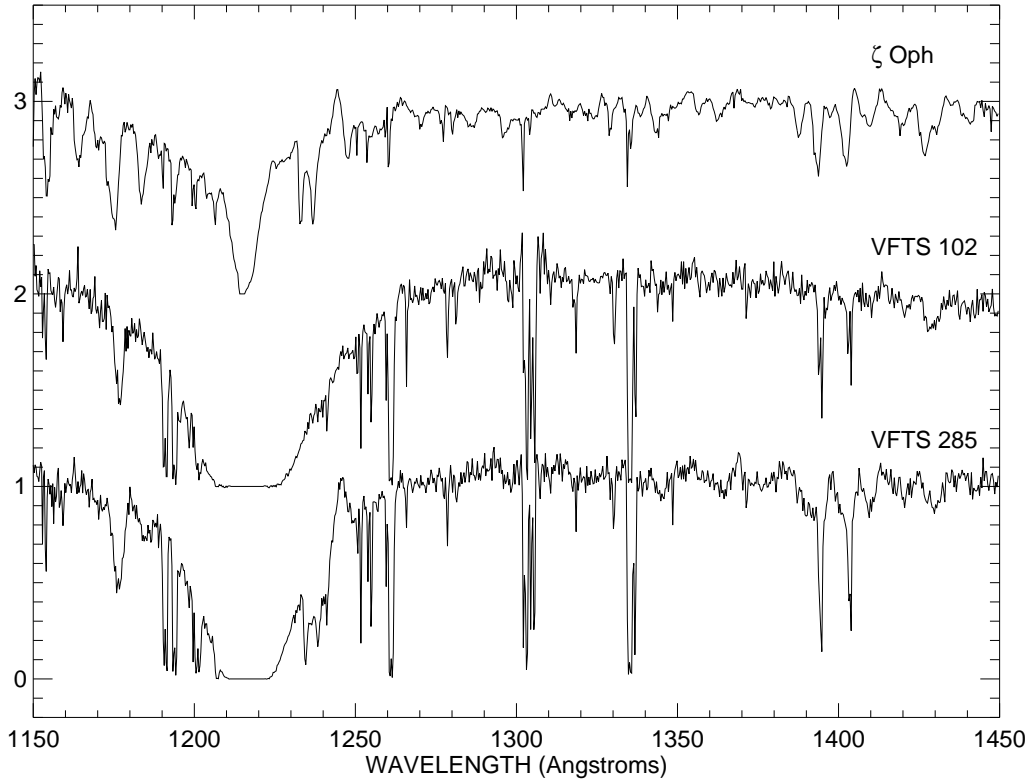}
\caption{Far-ultraviolet spectra of the three rapidly rotating stars. The spectra are normalized to a pseudo-continuum level of unity, and they are offset for clarity.  The broad, shallower features are formed in the photospheres, and the many sharp lines have an interstellar origin (as does the very broad Ly$\alpha$ $\lambda$ 1215 absorption line). 
}
\label{fig:fuvspectra}
\end{figure*}

\placefigure{fig:fuvspectra}

The IUE instrumentation suite consisted of two UV spectrographs, 
two apertures, two dispersion modes, and four cameras \citep{Boggess1978}.  Our spectra were 
obtained using the Short Wavelength Prime camera in the 
high resolution dispersion mode (Table~1).
For the purposes of this work, the spectra were flux normalized 
to unity in the relatively line-free regions, transformed 
to a $\log \lambda$ wavelength grid, and 
co-added to form one spectrum with a high $S/N$ ratio (Fig.~1). 

\subsection{Optical}

Our sample of optical spectra of VFTS~102 and VFTS~285 consists 
of new medium resolution spectra from the X-shooter spectrograph 
on the Very Large Telescope (VLT) plus archival 
spectra from the VLT Fibre Large Array Multi Element Spectrograph
(FLAMES) instrument used with the GIRAFFE spectrograph.
The optical spectra of $\zeta$~Oph were collected from the 
archive of the ESPaDOnS spectrograph
mounted on the Canada-France-Hawaii Telescope. 
The spectrograph properties are given in Table~1, 
and the averaged spectra appear in Figure~2.
  
\begin{figure*}
\centering
\includegraphics[angle=0,width=15cm]{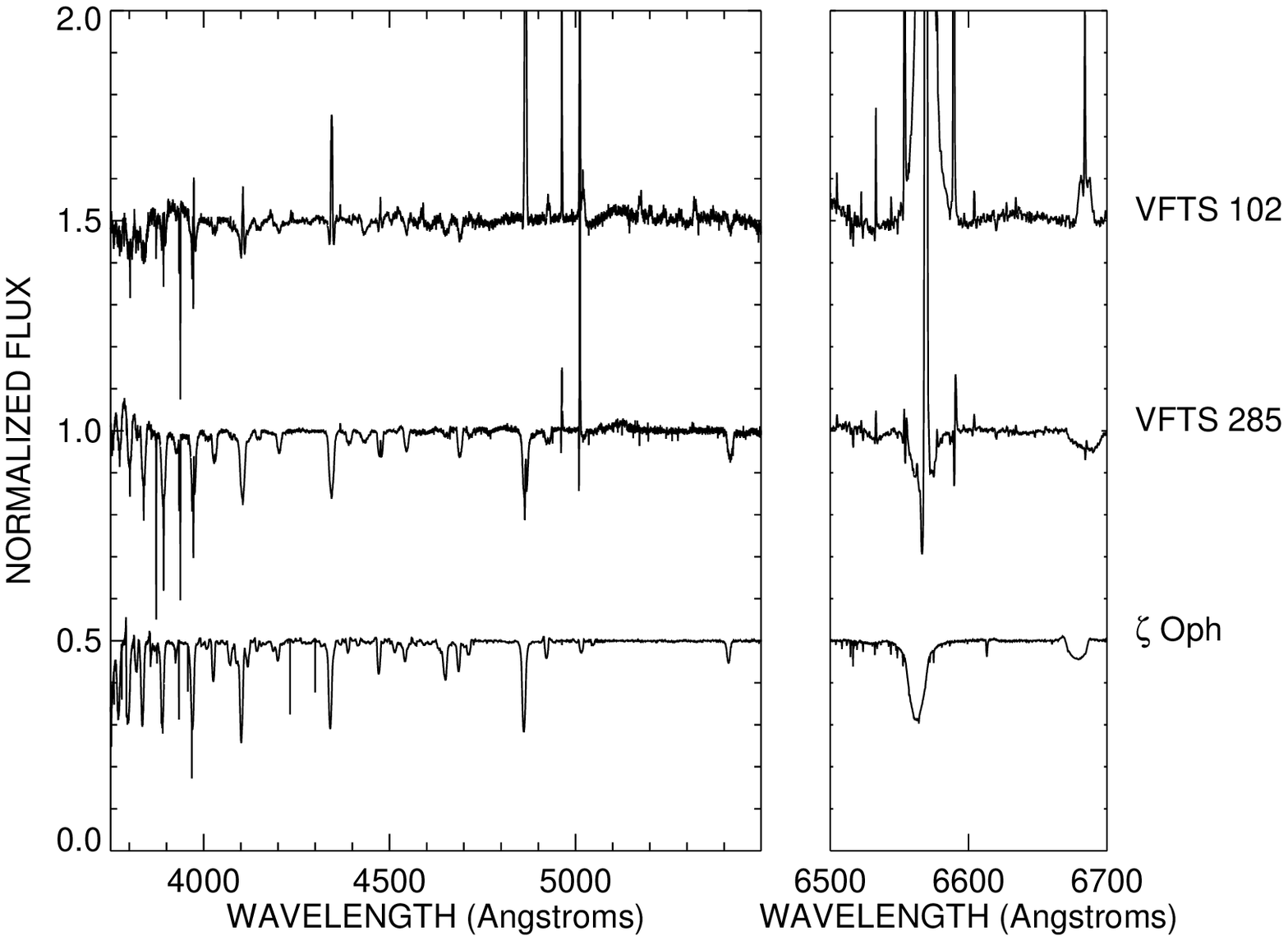}
\caption{The optical spectral region of the three stars
(from top to bottom) VFTS~102, VFTS~285, and $\zeta$ Oph. 
Most of the emission features are from background nebular emission or instrumental flaws, with the exception of the disk emission in the Balmer lines in the spectrum of VFTS~102 (especially H$\alpha$ $\lambda 6563$). 
}
\label{fig:threeOpt}
\end{figure*}

\placefigure{fig:threeOpt}
  
The X-shooter instrument was designed to record the spectra 
of a wide variety of astronomical objects, ranging from nearby 
faint point sources to bright extragalactic sources 
\citep{Vernet2011}. 
The observations for VFTS~102 were acquired under program
092.D-0108(A) (Przybilla),
and additional spectra were obtained for both VFTS~102 and VFTS~285
under program 098.D-0375(A) (Gies). 
The X-shooter instrument records spectra in three arms 
corresponding to increasing wavelength bands labelled 
as UVB, VIS, and NIR (see Table~1). 
Collectively, these spectra cover a
range from 3350 to 20800 \AA . 
The spectra were reduced by the standard pipeline, 
normalized to unity at the continuum, and then co-added 
with weighting factors determined by the $S/N$ ratio. 

The GIRAFFE spectrograph is a medium-to-high resolution spectrograph
that was designed to record the optical spectra of high spatial density
galactic and extragalactic objects \citep{Pasquini2002}. Spectral
observations for both VFTS~102 and VFTS~285 were obtained as 
part of the VLT-FLAMES Tarantula Survey of
massive stars in the 30 Doradus region in the LMC 
under programs 182.D-0222(A), (B), and (C). 
The observations were made
using the Medusa fibers on the GIRAFFE spectrograph, 
which allow for up to 132 objects to be observed at once. 
The fibers have an entrance aperture equal to $1.2$ arcsec on
the sky. The spectra utilized in this work cover a range from 3958 to
6820 \AA\ and are labelled as UV, VIS, and  NIR 
(corresponding to different bands than those 
with similar names for X-shooter; see Table~1). 
The reduced spectra were collected from the ESO Science 
Archive\footnote{http://archive.eso.org/wdb/wdb/adp/phase3\_spectral/form?phase3\_collection=GIRAFFE\_MEDUSA}, rectified to unity at the continuum, and co-added to form one high $S/N$ ratio spectrum.
In the final step, the X-shooter and GIRAFFE spectra were co-added
where possible on a uniform $\log\lambda$ grid for optimum $S/N$.
These coadded spectra were the focus of this study, and 
samples of the line profiles are presented in Figures 10 and 11 below.
We did not reduce the effective resolving power of the X-shooter 
spectra to match that of the GIRAFFE spectra 
because the rotational broadening of the stellar features far exceeds that of the 
instrumental broadening, and as a result the line profiles appeared identical in both
the X-shooter and GIRAFFE average spectra. 

ESPaDOnS is a high resolution spectrograph and
spectropolarimeter that was designed to record the optical spectrum
with a resolving power of
$R=\lambda/\triangle\lambda=68000$ \citep{Donati2003}. 
The ESPaDOnS spectra of $\zeta$~Oph were obtained from the
archive at the {\it PolarBase} website\footnote{http://polarbase.irap.omp.eu/} \citep{Petit2014}
as a part of the Magnetism in Massive Stars
({\it MiMeS}) survey that collected data from 2005 to 2013 for 560 O- and
B-type stars \citep{Wade2015,Wade2016}. 
The reduced spectra were rectified, and co-added on a 
uniform $\log\lambda$ wavelength grid. 

\subsection{Spectral Energy Distributions}

The rotation code described in \S4 calculates both the line profiles and the absolute monochromatic flux in the nearby continuum regions. 
The absolute flux predictions taken together with the distance and 
interstellar extinction can be compared to the observed spectral energy distribution (SED) to determine the stellar radius.
Figures 3, 4, and 5 show the observed SEDs for $\zeta$~Oph, VFTS~102, and VFTS~285, respectively, together with the model estimates for 16 wavelengths (\S4). 
The observed ultraviolet fluxes for comparison with the models were collected at the two FUV wavelengths from archival, high dispersion IUE spectra for $\zeta$~Oph and from the HST/COS spectra for VFTS~102 and VFTS~285.  The figures show instead low resolving power ($R=500)$ UV spectra from low dispersion IUE spectra for $\zeta$~Oph and rebinned versions of the HST/COS spectra for VFTS~102 and VFTS~285.  
The optical fluxes of $\zeta$~Oph are from the spectrophotometry of \citet{Burnashev1985}, and those for VFTS~102 and VFTS~285 are collected from various sources of broad-band photometry including 
VFTS \citep{Evans2011}, HTTP \citep{Sabbi2016}, 
SkyMapper \citep{Wolf2018}, and Gaia EDR3 \citep{Gaia2016,Gaia2021}.
The infrared fluxes are collected from SAGE \citep{Meixner2006}, 2MASS \citep{Skrutskie2006}, and WISE \citep{Wright2010}. 

Figures 3, 4, and 5 also show simple flux models for non-rotating stars from the TLUSTY grid \citep{Lanz2003} for the average temperature, gravity, abundance, 
distance, and extinction described in \S4.  
These model spectra have a low resolving power similar to the observed broad-band fluxes, and they serve to show the general SED trends with wavelength.  The SED of VFTS~102 displays an infrared excess from its circumstellar disk, and Figure~4 shows both the stellar component (dotted line) and combined stellar plus disk flux (solid line) for a simple power law expression for the disk flux (see \S4.2).  
We will adopt wavelength-interpolated estimates of the optical monochromatic fluxes below (\S4) directly from the observed values for the cases of $\zeta$~Oph and VFTS~285, and from the star plus disk model fit for the case of VFTS~102.

\begin{figure*}
\centering
\includegraphics[angle=-90,width=\textwidth]{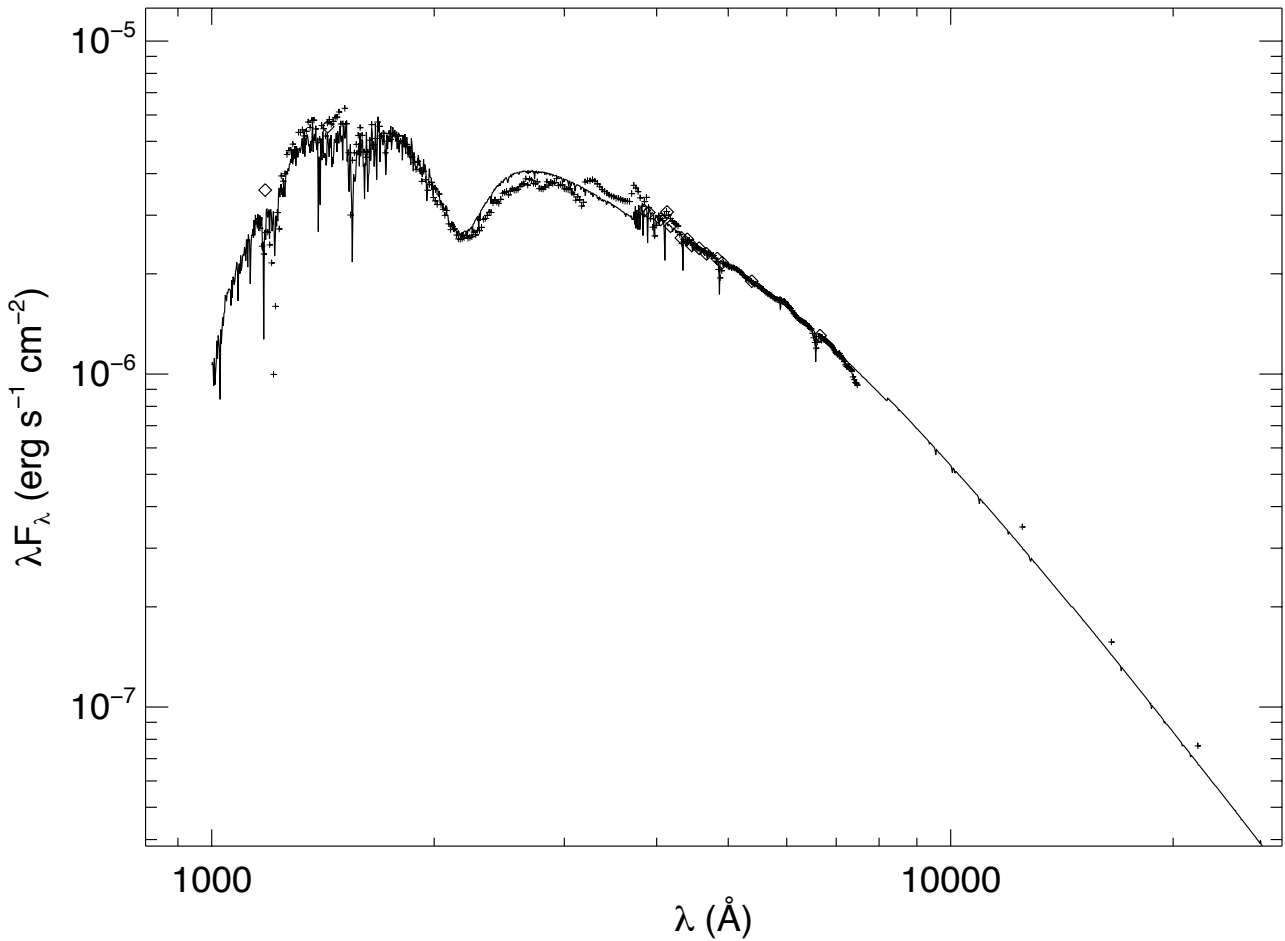}
\caption{
The observed spectral energy distribution (SED) of $\zeta$ Oph. The small crosses show the low resolving power observed fluxes, and the diamonds indicate the high resolving power fluxes calculated at 16 specific wavelengths using the rotation code (\S 4.2).  The solid line shows a low resolving power model SED from the TLUSTY code for a non-rotating star with the hemisphere-average temperature and gravity of the star (Table~4).  The model fluxes are attenuated for interstellar extinction using the reddening $E(B-V)$ given in Table~4.
}
\label{fig:sedzetaOph}
\end{figure*}

\placefigure{fig:sedzetaOph}

\begin{figure*}
\centering
\includegraphics[angle=-90,width=\textwidth]{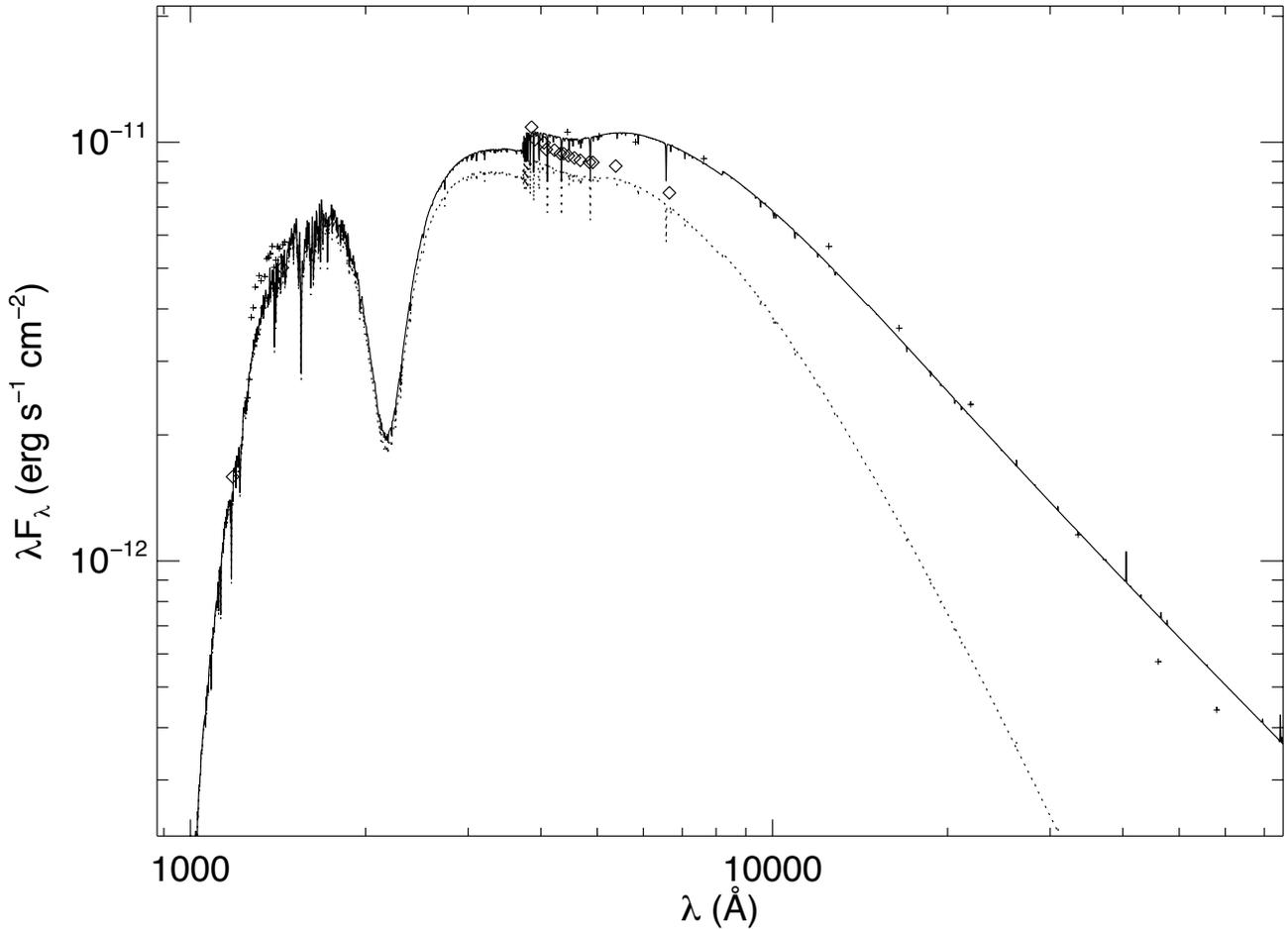}
\caption{
The observed spectral energy distribution (SED) of VFTS~102 in the same format as Fig.~3. The dotted line shows the SED of a TLUSTY model for the star alone, and the solid line presents the model of the combined flux of the star and its circumstellar disk.  
}
\label{fig:sedvfts102}
\end{figure*}

\placefigure{fig:sedvfts102}

\begin{figure*}
\centering
\includegraphics[angle=-90,width=\textwidth]{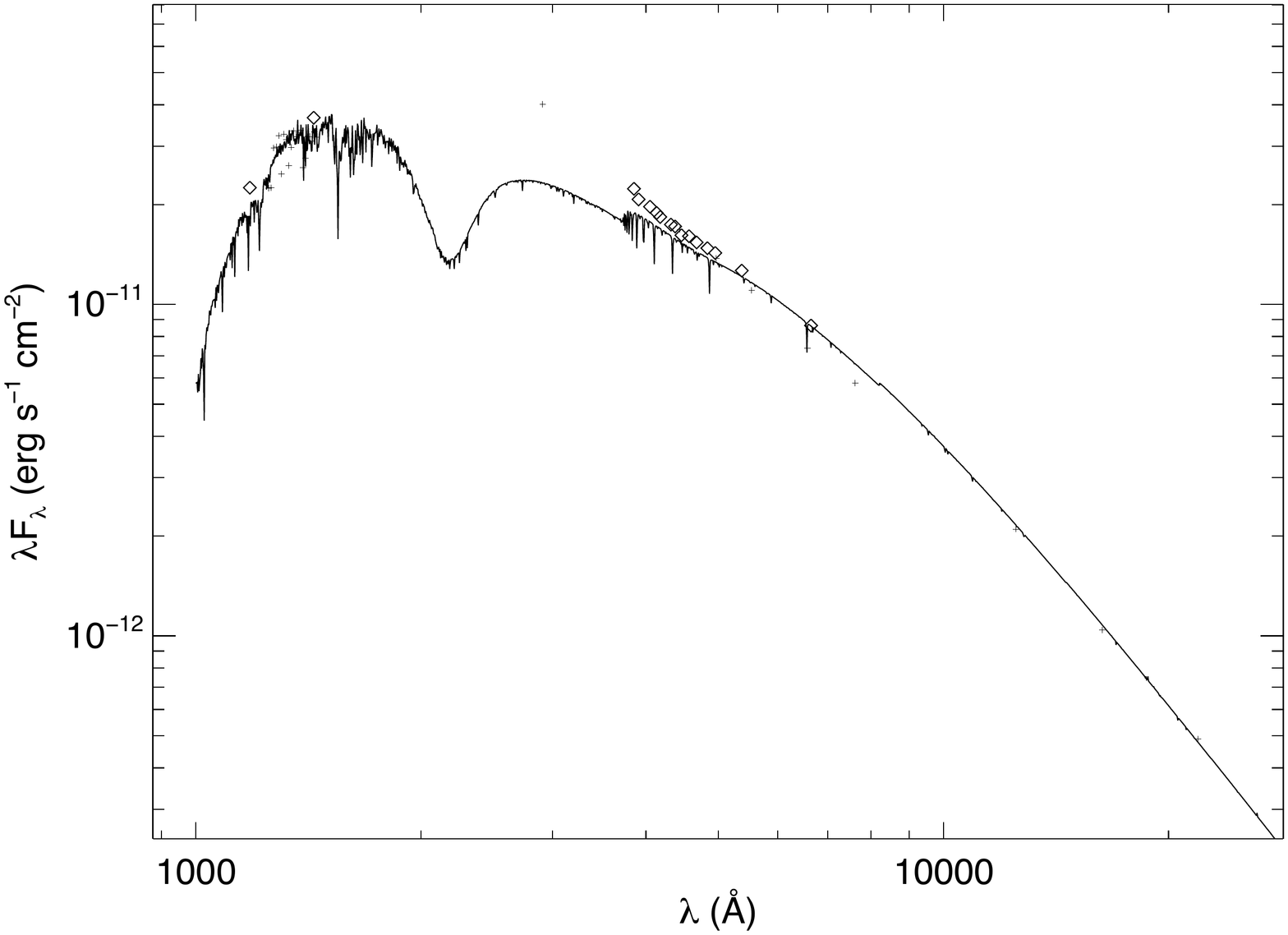}
\caption{
The observed spectral energy distribution (SED) of VFTS~285 in the same format as Fig. 3. 
}
\label{fig:sedvfts285}
\end{figure*}

\placefigure{fig:sedvfts285}

\section{Radial Velocities of VFTS~102 and VFTS ~285} 

The absorption lines in the spectra of both targets are extremely broad
and shallow. Consequently, we need the best $S/N$ ratio possible in order
to examine their rotationally broadened profiles. This can be
accomplished for the ground-based, optical spectra through co-addition
of the individual spectra.  However, we need to first check for any
evidence of radial velocity variability to insure that the spectra
are properly wavelength registered before co-addition and to explore the
possibility that these stars are spectroscopic binaries with faint
companions (the possible remnant donors of past mass transfer). We describe
below measurements of both the absorption and emission line features.
Our results are summarized in Table~2 (VFTS~102) and Table~3
(VFTS~285).

There are two main methods that we use to make our measurements,
the line bisector method for emission lines and the cross-correlation
method for absorption lines. The bisector method is useful for the
spectra of VFTS~102 which displays numerous emission lines 
formed in its circumstellar disk and in the surrounding nebula and SN remnant 
(known variously as 30 Dor B, \citealt{Chu1992}; 
N157B, \citealt{Chen2006}; and 
B0538-691, \citealt{Micelotta2009}).
The bisector method determines the line
center position by a Gaussian sampling of the line wings
\citep{Shafter1986}. 
This has the advantage of measuring emission wings from the disk
while ignoring the nebular emission that appears in the line core.
We form a template using two oppositely signed
Gaussians at offset positions from line center, and then the cross-correlation
function (CCF) is made using this template and the emission line
feature. The zero crossing of the resulting CCF yields the velocity
corresponding to the wing bisector position. The wings of emission
features form in the circumstellar disk close in to the star where the
rotating disk has the highest Keplerian orbital motion. Since the disk
is assumed to be centered on and tied to the star, by measuring the
disk’s radial velocity we are also measuring the star’s radial velocity.
In general we set the offset positions of the sampling Gaussians at
velocities where the emission declines to $25\%$ of the peak value.
However, if the feature has extraneous emission or absorption in the
center due to residual problems from nebular sky subtraction or disk
emission, we lower this threshold to around $5\%$ of the peak value in
order to avoid the central region. 

The second method was applied to the broad absorption lines in the
spectra of both stars. We generated a model spectrum which was rotationally broadened
to a projected rotational velocity of $v_{e} \sin i = 600$ km~s$^{-1}$
using the TLUSTY/SYNSPEC model flux spectra from the OSTAR2003 grid of
\citet{Lanz2003}. We then formed the CCF of the observed and model
spectra over a wavelength range encompassing specific absorption lines.
The exact regions that were used were limited to only the line profiles
and did not include the continuum between features.  The peak
position of the CCF yielded an estimate of radial velocity.


The results for VFTS~102 appear in Table~2 which lists the
heliocentric Julian date of mid-exposure, the spectrograph of origin,
and radial velocities from the emission and absorption lines. 
The radial velocity measurements for the X-shooter spectra 
are averages from the emission features (column 3) of 
H$\delta$, H$\beta$, H$\alpha$, 
\ion{H}{1} $\lambda\lambda 8502, 8545, 8598, 8665$, and
\ion{He}{1} $\lambda\lambda 5875, 6678, 7065$ 
and from the absorption features (column 4) of
H$\gamma$ and \ion{He}{2} $\lambda 4686$.  H$\gamma$ consists of a
broadened absorption feature with a double-peaked emission line in
the center. Despite the central emission, we were able to measure the
wings of the absorption line, so this radial velocity
is included among the absorption line average.

We attempted to measure the radial velocities of the absorption lines reported by
\citet{Dufton2011}, which consist of 
\ion{He}{1} $\lambda\lambda 4026, 4143, 4387$
and \ion{He}{2} $\lambda\lambda 4200, 4541, 4686$. 
However, we found that all but \ion{He}{2} $\lambda 4686$
were too broad and shallow for reliable measurements
in individual spectra.  \citet{Dufton2011} used 
co-added spectra to enable their measurements. 

The measurements of the GIRAFFE spectra include the emission features
H$\gamma$, H$\delta$, and H$\beta$. Several of the GIRAFFE spectra
have very low $S/N$ making the measurements difficult, and these are
excluded from consideration. 
All of the emission line measurements were made using the
line bisector method.  The only absorption line measured was 
\ion{He}{2} $\lambda 4686$.  In this case, we 
first cross-correlated the observed \ion{He}{2} $\lambda 4686$ profile with a
broadened model profile. We then measured the radial velocity using the
line bisector on the CCF.  This combination of methods allowed us
to obtain a measurement from the \ion{He}{2} $\lambda 4686$ feature 
despite its extremely shallow profile.
The measurement from the HST/COS spectrum was 
aquired by cross-correlating the observed spectrum with 
a rotationally broadened model spectrum.  
We used a short wavelength region that includes
\ion{N}{4} $\lambda 1168.6$, \ion{C}{3} $\lambda 1174.93$ (plus a
blend of six other features), and \ion{Si}{3} $\lambda 1178.012$
in addition to a mid-range region that includes \ion{Fe}{5} $\lambda\lambda 1370.303, 1370.947, 1371.217$ and \ion{Si}{4} $\lambda\lambda 1394, 1403$. The result is listed in column 4 of Table~2. 

The final rows of Table~2 list the error weighted averages of all the
measurements made from the X-shooter and
GIRAFFE spectra. The uncertainties are the standard
deviations of the individual measurements.
There is satisfactory agreement between the absorption and emission line
velocities, and this confirms that we are indeed measuring the radial
velocity of the star itself through measurements of the disk gas
emission.  
Additional verification comes from detailed fits made of the 
absorption lines presented in Section 5.2 below.  Each model line was shifted in wavelength space until it matched the observed profile. This yields a radial velocity for 
each absorption line in the mean spectrum, and the average for 
the nine lines used in the analysis yields a radial velocity 
of $262 \pm 8$ (standard deviation) km~s$^{-1}$, in agreement with the 
results presented in Table~2.
Furthermore, there is reasonable consistency in the 
results from X-shooter, GIRAFFE, and HST/COS.  
The weighted mean of the averages from both emission and absorption 
lines and all three instruments for VFTS~102 is $V_r = 267 \pm 3$ km~s$^{-1}$.

The velocity results from the absorption lines in the spectra of
VFTS~285 are given in Table~3. The radial velocities for the
X-shooter spectra were obtained by cross-correlating a series of
features with a rotationally broadened model. 
The features include H$\beta$, H$\gamma$, H$\delta$, and 
\ion{He}{2} $\lambda\lambda 4199, 4541, 4686, 5411$. 
The GIRAFFE measurements were obtained from cross-correlation
functions based upon 
H$\gamma$, H$\delta$, \ion{He}{1} $\lambda\lambda 3964, 4026, 4471$
and \ion{He}{2} $\lambda\lambda 4199,4541$. 
The measurement obtained for the HST/COS spectrum was obtained 
from the cross-correlation function of a short wavelength region
(including \ion{C}{3} $\lambda 1174.93$ and \ion{Si}{3} $\lambda 1178.0$) and a mid-wavelength region (\ion{Fe}{5}
$\lambda\lambda 1370.303, 1370.947, 1371.217$). 
The bottom rows of Table~3 show that there is good agreement 
between the averages from X-shooter and GIRAFFE. 
The mean of the X-shooter, GIRAFFE, and HST/COS results 
is $V_r = 250 \pm 6$ km~s$^{-1}$.

\placetable{tab2}      
\begin{deluxetable*}{cccc}
\tabletypesize{\scriptsize}
\movetableright=0.1mm
\tablenum{2}
\tablecaption{Radial Velocity Measurements: VFTS~102\label{tab2}}
\tablewidth{0pt}
\tablehead{
\colhead{Heliocentric} & 
\colhead{Spectrograph} &
\colhead{$V_r$(emission)} &
\colhead{$V_r$(absorption)} \\
\colhead{Julian Date} & 
\colhead{Name} &
\colhead{(km s$^{-1}$)} &
\colhead{(km s$^{-1}$)} 
}
\startdata
2454822.754  &   GIRAFFE      & 265 $\pm$ 18 &     \nodata       \\
2454822.797  &   GIRAFFE      & 268 $\pm$ 10 &     \nodata       \\
2454822.840  &   GIRAFFE      & 266 $\pm$ 11 &     \nodata       \\
2454825.740  &   GIRAFFE      & 269 $\pm$ 15 &    252 $\pm$ 27   \\
2454825.783  &   GIRAFFE      & 268 $\pm$ 16 &    235 $\pm$ 11   \\
2454825.826  &   GIRAFFE      & 267 $\pm$ 18 &    248 $\pm$ 4    \\ 
2454858.720  &   GIRAFFE      & 266 $\pm$ 8  &     \nodata       \\
2454889.552  &   GIRAFFE      & 262 $\pm$ 9  &     \nodata       \\
2455114.772  &   GIRAFFE      & 199 $\pm$ 30 &     \nodata       \\ 
2456605.711  &   X-shooter    & 271 $\pm$ 2  &    268 $\pm$ 11   \\ 
2456606.726  &   X-shooter    & 274 $\pm$ 2  &    267 $\pm$ 10   \\
2457299.761  &   GIRAFFE      & 274 $\pm$ 18 &     \nodata       \\
2457335.798  &   GIRAFFE      & 264 $\pm$ 10 &     \nodata       \\
2457339.714  &   GIRAFFE      & 261 $\pm$ 12 &     \nodata       \\
2457339.796  &   GIRAFFE      & 263 $\pm$ 17 &     \nodata       \\
2457366.740  &   GIRAFFE      & 260 $\pm$ 16 &     \nodata       \\
2457394.668  &   GIRAFFE      & 266 $\pm$ 9  &     \nodata       \\
2457398.739  &   GIRAFFE      & 262 $\pm$ 12 &     \nodata       \\
2457400.582  &   GIRAFFE      & 251 $\pm$ 40 &     \nodata       \\
2457402.618  &   GIRAFFE      & 265 $\pm$ 19 &     \nodata       \\
2457415.563  &   GIRAFFE      & 262 $\pm$ 12 &     \nodata       \\
2457416.575  &   GIRAFFE      & 261 $\pm$ 14 &     \nodata       \\
2457417.579  &   GIRAFFE      & 271 $\pm$ 11 &     \nodata       \\
2457418.628  &   GIRAFFE      & 273 $\pm$ 14 &     \nodata       \\
2457420.638  &   GIRAFFE      & 268 $\pm$ 11 &     \nodata       \\
2457423.615  &   GIRAFFE      & 266 $\pm$ 16 &     \nodata       \\
2457427.623  &   GIRAFFE      & 269 $\pm$ 9  &     \nodata       \\
2457432.533  &   GIRAFFE      & 280 $\pm$ 13 &     \nodata       \\
2457622.882  &   GIRAFFE      & 267 $\pm$ 14 &     \nodata       \\
2457681.726  &   GIRAFFE      & 272 $\pm$ 9  &     \nodata       \\
2457692.759  &   GIRAFFE      & 267 $\pm$ 13 &     \nodata       \\
2457698.764  &   GIRAFFE      & 258 $\pm$ 10 &     \nodata       \\
2457726.726  &   GIRAFFE      & 264 $\pm$ 12 &     \nodata       \\
2457755.380  &   HST/COS      & \nodata      &    223 $\pm$ 15   \\
2457797.584  &   X-shooter    & 267 $\pm$ 2  &    269 $\pm$ 11   \\
2457819.533  &   X-shooter    & 268 $\pm$ 2  &    257 $\pm$ 11   \\
\tableline
Average      &   X-shooter    & 270 $\pm$ 3  &    265 $\pm$ 6    \\
Average      &   GIRAFFE      & 266 $\pm$ 13 &    247 $\pm$ 9    \\
\enddata
\end{deluxetable*}

\newpage
  
\placetable{tab3}      
\begin{deluxetable*}{ccc}
\tabletypesize{\scriptsize}
\tablenum{3}
\tablecaption{Radial Velocity Measurements: VFTS~285\label{tab3}}
\tablewidth{0pt}
\tablehead{
\colhead{Heliocentric} & 
\colhead{Spectrograph} & 
\colhead{$V_r$(absorption)} \\
\colhead{Julian Date} & 
\colhead{Name} & 
\colhead{(km s$^{-1}$)}
}
\startdata
2454794.686  &    GIRAFFE   &  248 $\pm$ 4   \\
2454794.730  &    GIRAFFE   &  248 $\pm$ 4   \\
2454794.828  &    GIRAFFE   &  270 $\pm$ 4   \\
2454798.803  &    GIRAFFE   &  276 $\pm$ 5   \\
2454836.645  &    GIRAFFE   &  245 $\pm$ 4   \\
2454836.691  &    GIRAFFE   &  245 $\pm$ 4   \\
2454867.558  &    GIRAFFE   &  251 $\pm$ 4   \\
2457299.761  &    GIRAFFE   &  232 $\pm$ 4   \\
2457332.749  &    GIRAFFE   &  261 $\pm$ 4   \\
2457335.798  &    GIRAFFE   &  251 $\pm$ 4   \\
2457339.714  &    GIRAFFE   &  258 $\pm$ 4   \\
2457339.796  &    GIRAFFE   &  255 $\pm$ 4   \\
2457366.740  &    GIRAFFE   &  238 $\pm$ 4   \\
2457379.567  &    GIRAFFE   &  269 $\pm$ 4   \\
2457394.668  &    GIRAFFE   &  243 $\pm$ 4   \\
2457398.739  &    GIRAFFE   &  249 $\pm$ 4   \\
2457400.582  &    GIRAFFE   &  268 $\pm$ 5   \\ 
2457402.618  &    GIRAFFE   &  229 $\pm$ 12  \\
2457415.563  &    GIRAFFE   &  254 $\pm$ 4   \\
2457416.575  &    GIRAFFE   &  268 $\pm$ 4   \\
2457417.579  &    GIRAFFE   &  268 $\pm$ 4   \\
2457418.628  &    GIRAFFE   &  258 $\pm$ 4   \\
2457420.638  &    GIRAFFE   &  234 $\pm$ 4   \\
2457421.711  &    GIRAFFE   &  270 $\pm$ 3   \\
2457423.615  &    GIRAFFE   &  253 $\pm$ 4   \\
2457427.623  &    GIRAFFE   &  267 $\pm$ 4   \\
2457432.533  &    GIRAFFE   &  254 $\pm$ 4   \\
2457488.626  &    HST/COS   &  252 $\pm$ 17	 \\
2457622.882  &    GIRAFFE   &  247 $\pm$ 5   \\
2457651.777  &    GIRAFFE   &  233 $\pm$ 6   \\
2457681.726  &    GIRAFFE   &  263 $\pm$ 4   \\
2457692.759  &    GIRAFFE   &  268 $\pm$ 4   \\
2457698.764  &    GIRAFFE   &  254 $\pm$ 4   \\
2457724.723  &    GIRAFFE   &  243 $\pm$ 9   \\
2457726.726  &    GIRAFFE   &  263 $\pm$ 4   \\
2457765.726  &   X-shooter  &  240 $\pm$ 3   \\ 
2457766.733  &   X-shooter  &  241 $\pm$ 3   \\
2457775.634  &   X-shooter  &  256 $\pm$ 5   \\
2457787.536  &   X-shooter  &  243 $\pm$ 4   \\
2457790.533  &   X-shooter  &  249 $\pm$ 4   \\
2457790.570  &   X-shooter  &  253 $\pm$ 3   \\
\tableline
Average      &   X-shooter  & 247 $\pm$ 7    \\
Average      &    GIRAFFE   & 259 $\pm$ 13   \\ 
\enddata
\end{deluxetable*}

The standard deviation between observations (external error) 
is approximately the same as the mean of the individual 
error estimates (internal error) for both stars, 
so they appear to be radial velocity constant. 
Thus, we can reasonably perform a simple co-addition of the 
spectra to improve the $S/N$ without needing to consider 
shifting individual spectra to account for any binary orbital motion.
The final mean velocities are $267 \pm 3$ and $250 \pm 6$ km~s$^{-1}$
for VFTS~102 and VFTS~285, respectively, which are the error weighted
mean and uncertainty from the sample averages at the bottom of 
Tables 2 and 3.  These radial velocities are comparable 
to the average for the single, B-type stars in the 
30~Dor region, $272 \pm 12$ km~s$^{-1}$, found by 
\citet{Evans2015}. In both cases, our results are 
slightly higher than found in earlier work: 
$228 \pm 6$ km~s$^{-1}$ \citep{Dufton2011} and 
$225 \pm 11$ km~s$^{-1}$ \citep{Sana2013} for VFTS~102
and $230 \pm 4$ km~s$^{-1}$ \citep{Sana2013} for VFTS~285. 
However, given the difficulty of measuring such 
broad and shallow absorption lines and differences 
in our methods, we doubt that the resulting differences 
in velocity are significant.  

\section{Spectral Synthesis Models} 

Our primary goal is to compare models of the flux emitted by very
rapidly rotating stars with observed spectral line profiles and the
associated spectral energy distribution. Each model is based on parameters for its equatorial rotational velocity,
physical parameters, and axial orientation to our line of sight. Fits of
the model spectra to the observed spectra help inform final estimates of
all these parameters. In this section, we describe the elements of the
model and the parameter fitting methods. The results of the model
fitting are discussed in \S5.  

\subsection{Method}  

We utilize a numerical code that simulates the distorted shape and
latitude-dependent photospheric properties of a rapidly rotating star
that is viewed at an inclination angle $i$ between the axis of rotation
and the line of sight. The spectral line synthesis code is written in
IDL, and the original version was presented by \citet{Huang2006}. The
shape of the star is defined by Roche geometry which assumes that most of
the mass is concentrated towards the stellar core. The stellar
photosphere is represented by a grid of 40,000 surface elements with
approximately equal area that are distributed in co-latitude $\theta$
and azimuth $\phi$. An integration is made of the flux from each surface
element by first calculating the angle between the surface normal and
the line of sight in order to determine if the element is situated on
the visible hemisphere of the star. For each element, the code
determines a local effective temperature (dependent on the gravity
darkening model; \S4.4), effective gravity (gravitational plus
centrifugal), area of the surface element projected in the sky
(dependent on $\mu$ = cosine of the angle between the surface normal and
line of sight), and the rotational radial velocity (assuming solid body
rotation). These parameters are used to interpolate in a pre-computed
grid of spectral specific intensities as a function of wavelength, $\mu$,
$T_{\rm eff}$, and $\log g$ associated with an assumed chemical
abundance. The product of the projected area and the specific intensity
yields the flux increment from the element, and the sum of all such
increments gives the total flux to be compared with the observations. 
The outputs are a wavelength dependent line profile and a monochromatic
flux estimate for the immediate vicinity of a particular spectral
feature or line blend. This type of spectrum synthesis is common among
many past investigations of rotational shape and spectral line
broadening \citep{Collins1963,Stoeckley1968,Howarth2001,Townsend2004,Aufdenberg2006,Abdul-Masih2020}.
 
The model parameters are listed in Table~4 together with the derived
values for the three target stars of this study. Column 2 identifies
those parameters that are fit (F), set in advance (S), and derived in
the model from the previous parameters (D). The set of fitting
parameters includes the projected rotational velocity $v_{e} \sin i$,
rotational axis inclination $i$, polar radius $R_p$, stellar mass $M$,
polar effective temperature $T_p$, He abundance $y$ by number relative to H, and interstellar reddening $E(B-V)$. The set parameters were
determined by independent constraints, and they include the adopted
gravity darkening law (\S4.4), distance $d$, and ratio of total-to-selective
extinction $R_V$ (\S4.2). The remaining derived parameters
describe the physical characteristics of the star and the
goodness-of-fit of the model. The derived rotation parameters are the
equatorial rotational velocity $v_e$, the critical velocity in the Roche
model $v_c$ (where the equatorial radius is $1.5\times$ the polar radius
and centrifugal acceleration balances gravity at the equator), and the
ratio of the angular velocity to the critical angular velocity $\Omega /
{\Omega}_c$ \citep{Rieutord2016}. The other physical parameters for the
star include the equatorial radius $R_e$, the logarithm of the effective
gravity at the pole $\log g_p$, at the equator $\log g_e$, averaged
over the visible hemisphere $\log g$(avg), the effective temperature at
the equator $T_e$, and the flux-weighted average temperature $<T>$(avg).
We also calculate the area-integrated average temperature $<T>$(all) and 
logarithm of the total luminosity $\log L/L_\odot$ by integrating over the
entire surface.  Those parameters listed with solar units were obtained
by adopting the IAU recommended nominal values for the Sun
\citep{Prsa2016}.  

\begin{deluxetable}{lcccccc}
\label{tab:results}
\tabletypesize{\scriptsize}
\tablenum{4}
\tablecaption{Summary of Rotational Parameters\label{tab4}}
\tablewidth{0pt}
\tablehead{
\colhead{Parameter} & 
\colhead{Kind\tablenotemark{a}} & 
\colhead{$\zeta$ Oph} & 
\colhead{$\zeta$ Oph\tablenotemark{b}} & 
\colhead{VFTS 102} & 
\colhead{VFTS 285} & 
\colhead{VFTS 285} 
}

\startdata
Gravity darkening          & S & $\omega$-model    & von Zeipel        & $\omega$-model    & $\omega$-model    & von Zeipel        \\
$v_{e}\sin i$ (km s$^{-1}$)    & F & $383 \pm 33$      & 400               & $649 \pm 52$      & $610 \pm 41$      & $629 \pm 34$      \\
$i$ (degrees)              & F & $72 \pm 17$       & $70 \pm 10$       & $89 \pm 13$       & $71 \pm 18$       & $60 \pm 12$       \\
$v_e$ (km s$^{-1}$)        & D & 413               & 425               & 649               & 648               & 726               \\
$v_c$ (km s$^{-1}$)        & D & 508               & \nodata           & 652               & 796               & 795               \\
${\Omega} / {\Omega}_c$    & D & 0.95              & $0.90 \pm 0.03$   & 1.00              & 0.95              & 0.99              \\
$R_p$ ($R_\odot$)          & F & $6.32 \pm 0.84$   & 7.5               & $5.41 \pm 1.55$   & $5.58 \pm 0.39$   & $5.53 \pm 0.39$   \\
$R_e$ ($R_\odot$)          & D & 8.11              & \nodata           & 8.08              & 7.16              & 7.65              \\
$M$ ($M_\odot$)            & F & $12.8 \pm 3.6$    & 20.0              & $18 \pm 6$        & $28 \pm 8$        & $27 \pm 3$        \\
$\log g_p$ (dex cgs)       & D & 3.94              & $3.99 \pm 0.05$   & 4.23              & 4.39              & 4.39              \\
$\log g_e$ (dex cgs)       & D & 3.35              & 3.58              & 2.26              & 3.81              & 3.47              \\
$\log g$(avg) (dex cgs)    & D & 3.62              &                   & 3.64              & 4.05              & 4.01              \\
$T_p$ (kK)                 & F & $36.0 \pm 1.6$    & $39.0 \pm 1.0$    & $40.1 \pm 2.8$    & $40.2 \pm 2.7$    & $43.0 \pm 3.2$    \\
$T_e$ (kK)                 & D & 28.3              & 30.7              & 23.2              & 31.6              & 25.3              \\
$<T>$(avg) (kK)            & D & 31.4              & \nodata           & 32.9              & 34.9              & 35.0              \\
$<T>$(all) (kK)            & D & 31.4              & 34.3              & 33.3              & 35.0              & 33.3              \\
$\log (L/L_\odot)$         & D & 4.67              & 4.96              & 4.75              & 4.76              & 4.71              \\
$y=N{\rm (He)}/N{\rm (H)}$ & F & $0.24 \pm 0.07$   & $0.20 \pm 0.03$   & $0.20 \pm 0.16$   & $0.34 \pm 0.14$   & $>0.4$            \\
$d$ (kpc)                  & S & $0.139 \pm 0.016$ & 0.14              & $49.6 \pm 0.6$    & $49.6 \pm 0.6$    & $49.6 \pm 0.6$    \\
$R_V$                      & S & $2.55 \pm 0.24$   & \nodata           & 2.76              & 2.76              & 2.76              \\
$E(B-V)$ (mag)             & F & $0.28 \pm 0.04$   & \nodata           & $0.46 \pm 0.07$   & $0.28 \pm 0.03$   & $0.27 \pm 0.03$   \\
$\chi^2_\nu$ [line]        & D & 16.5              & \nodata           & 1.7               & 4.1               & 4.1               \\
$\sigma$ ($F_o/F_m$)       & D & 0.042             & \nodata           & 0.127             & 0.056             & 0.057             \\
\enddata
  
\tablenotetext{a}{D = derived from model; F = fit; S = set.}
\tablenotetext{b}{From \citet{Howarth2001}.}
 
\end{deluxetable}

There are several simplifications in the model that are justifiable
assumptions. The shape of a star is subject to differential rotation,
but detailed calculations suggest that differential rotation is modest
in rapidly rotating massive stars (varying by only a few percent with
colatitude; \citealt{EspinosaLara2013}). Consequently we expect that our
neglect of differential rotation and the use of the Roche model for the
stellar surface are good approximations \citep{Zahn2010,Rieutord2016}. 
The spectral specific intensity calculations are based upon
plane-parallel stellar atmospheres (\S4.3), so effects due to
extended atmospheres and stellar winds are not treated. However, we
expect any such effects to be minimal in the case of the three targets
discussed here. We showed in an earlier paper \citep{Shepard2020} that
wind features are present in some spectral lines in the ultraviolet, and
we discuss the influence of the circumstellar disk of VFTS~102 below
(\S4.2). Finally, we are neglecting any processes related to
macroturbulence in this analysis. \citet{Simon-Diaz2007} used a Fourier
transform method to analyze the broadened spectral line profiles of
O-type stars to extract both the projected rotational velocity and the
macroturbulent velocity, and they argue that macroturbulence generally
becomes a significant contributor to line broadening among the more
luminous supergiant stars \citep{Simon-Diaz2014}. The net line
broadening varies approximately as the quadratic sum of the rotational
and macroturbulent velocities, and because the rotational component is
so dominant in the stars discussed here, we can safely neglect any 
macroturbulent broadening terms. 

\subsection{Integrated Flux} 

The model calculates the integrated monochromatic flux produced by the
star, and this can be compared to the observed flux to help estimate the
stellar radius. For a spherical star, the ratio of the observed to
emitted flux is $$F({\rm obs}) / F({\rm em}) = {1\over 4} \theta^2
10^{-0.4 A_\lambda} $$ where $\theta$ is the angular diameter and
$A_\lambda$ is the wavelength-dependent extinction. Thus, this ratio
becomes an important criterion to establish the stellar polar radius of
a rotating star $R_p$ once the distance and extinction are known. Here
we fix the known distances of the targets and solve for a reddening
$E(B-V)$ that sets the amount of interstellar extinction according to an
adopted value of the ratio of total-to-selective extinction $R_V$.

Our adopted distances are listed in Table~4. The distance of
$\zeta$~Oph is taken from the parallax measurement from Gaia EDR3
\citep{BailerJones2021}, and this value is in good agreement with other
independent estimates \citep{Gordon2018}. The distance of VFTS~102 and
VFTS~285 is set to the accurate LMC distance from
\citet{Pietrzynski2019} based upon eclipsing binaries and other standard
candles in the LMC. 

The wavelength dependent extinction curve is taken from the model 
described by \citet{Fitzpatrick1999} that is determined from the 
reddening $E(B-V)$ and ratio of total-to-selective extinction $R_V$.
We solve for $E(B-V)$ by comparing the observed and model SED 
using an adopted value for $R_V$ (Table~4). This is set to the 
result from \citet{Zuo2021} for $\zeta$~Oph. Unfortunately, 
there are no published results on $R_V$ for the two LMC targets.  
We adopted the value of $R_V = 2.76 \pm 0.09$ for the nearby 
LMC2 supershell region determined by \citet{Gordon2003} (and we used 
their special extinction prescription for VFTS~102 and VFTS~285),
although we caution that the actual value may vary considerably 
among the stars of the 30~Dor region.  For example, both 
\citet{Maiz-Apellaniz2014} and \citet{DeMarchi2019} find a 
larger value of the ratio, $R_V \approx 4.5$.

Special care is required in the flux analysis of VFTS~102, because this
star has an extensive circumstellar disk that also contributes to the 
observed flux. The SED of VFTS~102 (Fig.~4) shows a strong infrared 
excess that is a common signature of circumstellar disks around Be stars
\citep{Vieira2015,Klement2019}. The disk emission is often represented 
approximately as a power law \citep{Waters1986}, 
$$F^{\rm tot}_\lambda /
F^\star_\lambda = (1 + c_d (\lambda / \lambda_0)^x) ~10^{-0.4
A_\lambda}$$ 
where $F^{\rm tot}_\lambda$ is the observed star plus disk
flux, $F^\star_\lambda$ is the stellar flux rescaled by distance, 
$\lambda$ is wavelength in reference to a standard wavelength $\lambda_0 =
1$ $\mu$m, $A_\lambda$ is the extinction, and $c_d$ and $x$ are the
power law parameters describing the infrared excess.  A simple fit of
the SED using a TLUSTY model for the stellar flux was obtained with $c_d
= 0.78$ and $x=1.6$, and is shown with the observed SED in Figure~4.  

This disk flux excess in the spectrum of VFTS~102 has two important
consequences: (1) the model stellar flux must be increased 
by the amount of the flux excess before comparison with the observed
flux, and (2) the model line depths need to be reduced to account for
the excess continuum flux (sometimes referred to as ``line veiling''). 
We did this by calculating a line depth factor $l$ at the central
wavelength of each line, $$l = F^\star / (F^\star + F^{\rm disk}) = 1 /
(1 + c_d (\lambda / \lambda_0)^x).$$ The model stellar fluxes were
divided by $l$ to rescale them to the total star plus disk flux, and the
continuum normalized line depths were multiplied by $l$ to account for
the added disk continuum, i.e., $$s({\rm disk~corrected}) = l \times
s({\rm model}) + 1 - l$$ where $s$ is the continuum normalized spectrum
(\S4.5). This approach corrects for the added continuum flux of
the disk, but not for any stellar flux that
may be obscured by the disk if seen edge on. 
The calculations necessary to account for disk obscuration are outside the scope of this paper, and we therefore leave this complication to future work. 

\subsection{Specific Intensity Profiles} 

The core of the simulation is the radiative specific intensity that 
we derive from two TLUSTY grids of line-blanketed, non-LTE, model
atmospheres, OSTAR2002 \citep{Lanz2003} and BSTAR2006 \citep{Lanz2007}.
High resolving power model spectra are calculated from these model 
atmospheres using the SYNSPEC radiative transfer code \citep{Hubeny2017}.
These assume the solar abundance pattern from \citet{Grevesse1998} 
for Galactic ``G'' models (with a He abundance by number relative to 
H of $y=0.10$) that we use in the line synthesis for $\zeta$~Oph. 
We adopt their LMC ``L'' models that have the same H and He abundance 
with all other elements reduced by half \citep{Rolleston2002} for 
the analyses of VFTS~102 and VFTS~285.  We found it necessary to 
explore models with greater than solar He abundance (\S4.7), 
and this was done by a numerical up-scaling of the He to H number ratio by 
$2\times$, $3\times$, and $4\times$ the solar value with the 
SYNSPEC code.  We caution that this is an approximation that is 
not fully self-consistent with the abundances assumed in the 
TLUSTY atmospheres.   

The BSTAR2006 grid covers the temperature range from 15 to 30 kK
for an assumed microturbulent velocity of 2 km~s$^{-1}$, while 
the OSTAR2003 grid ranges from 27.7 to 55 kK using a microturbulent 
velocity of 10 km~s$^{-1}$.  In order to create a smooth transition 
between these differing cases, we formed a temperature grid with 
a step size of 1 kK (like that in the BSTAR2006 grid), and then 
interpolated in the temperature overlap region scaling by 
$75/25\% $, $50/50\% $, and $25/75\% $ between the BSTAR2006/OSTAR2003
grids at 28, 29, and 30 kK, respectively. 

For each spectral region, the SYNSPEC results were transformed to 
a common and equally spaced wavelength grid to create a specific intensity
matrix, $I(\lambda, \mu, T_{\rm eff}, \log g)$ for 10 equal steps in 
$\mu$, the cosine of the angle between the normal and line of sight, 
from 0.1 to 1.0, 41 steps in $T_{\rm eff}$ from 15 to 55 kK, and 
12 steps in $\log g$ from 2.0 to 4.75.  These matrices were calculated 
for two regions in the far-ultraviolet and 14 regions in the optical 
spectrum for comparison with the observed spectra (\S5). 
We show an example of these specific intensity profiles in Figure 6. 
These show that the central depths of strong lines are almost the 
same at all $\mu$ angles (formed high in the atmosphere where the temperature and line source function are approximately constant) while the 
continuum levels decline from $\mu=1$ to 0.1 (limb darkening associated 
with the drop in temperature and source function higher in the atmosphere). 
Thus, in a continuum normalized representation, the spectral line 
depths look relatively weaker at the limb ($\mu = 0$) compared to 
the center of the stellar disk ($\mu = 1$).  This is a reminder that 
the observed rotational broadening is not strictly a convolution of 
a fixed depth photospheric profile and a rotational broadening function. 

\begin{figure*}
\includegraphics[angle=90,width=\textwidth]{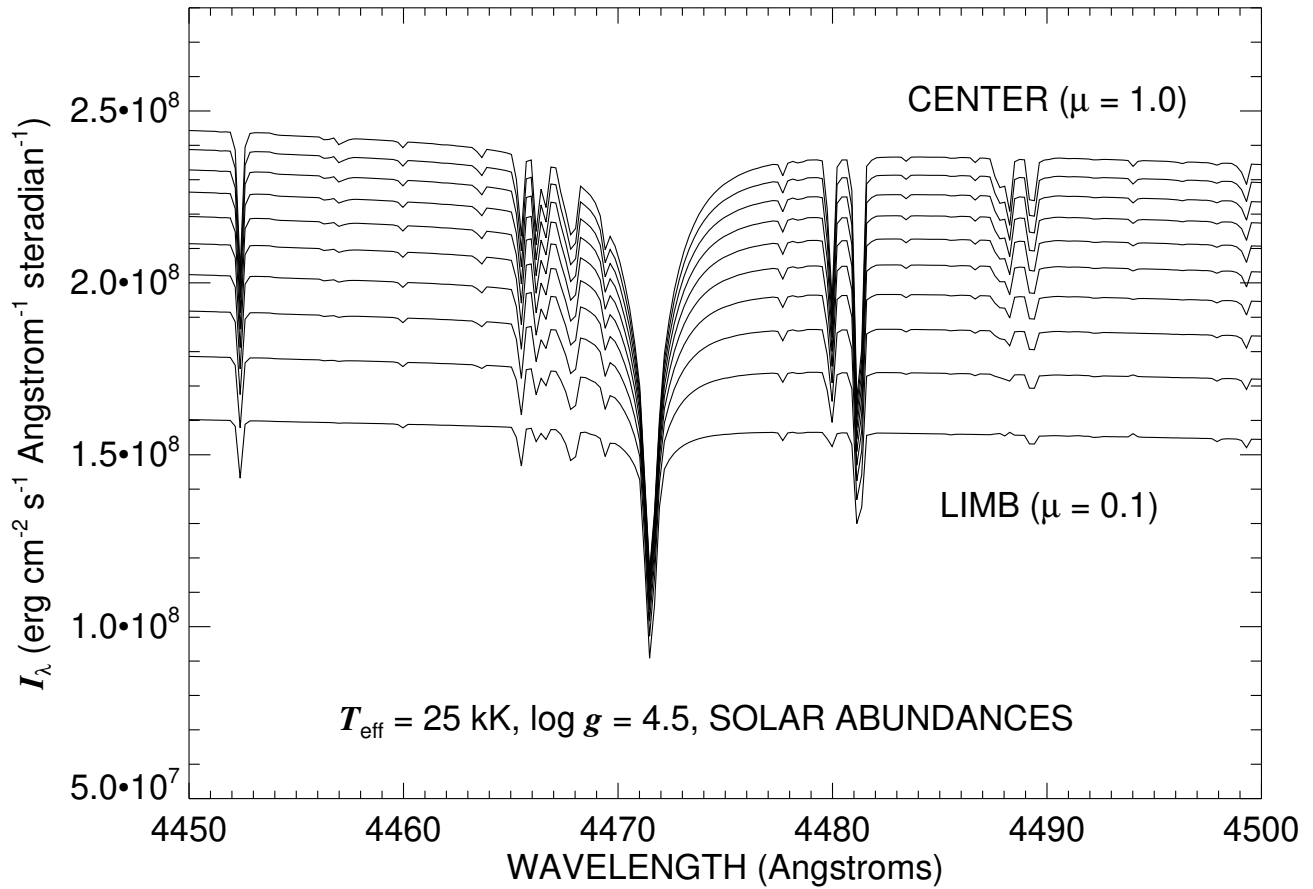}
\caption{An example of the model specific intensity profiles for the 
region in the vicinity of \ion{He}{1} $\lambda 4471$.  The plots 
show $I_\lambda$ for $\mu =0.1$ to $\mu=1.0$ at steps of 
$\triangle\mu =0.1$ from bottom to top.  Continuum limb darkening 
is evident (darker at the limb) while the central line depth of 
\ion{He}{1} $\lambda 4471$ is relatively constant.  
}
\label{fig:specint}
\end{figure*}

\placefigure{fig:specint}

Our analysis of the spectral lines is based upon this 
framework of specific intensities from the published TLUSTY 
grids, so the results are dependent on the approximations 
in the code in addition to those described above.  For example, 
TLUSTY includes the turbulent pressure associated with 
microturbulence as a component of the total pressure.  Thus, 
the pressure treatment is different between the BSTAR2006
and OSTAR2003 grids because of the different assumed values 
of microturbulence (2 and 10 km~s$^{-1}$, respectively), 
and this might influence the model Stark broadening of the 
H Balmer lines (important in the derivation of stellar mass;
Section 4.6).  However, the BSTAR2006 grid does include the 
larger 10 km~s$^{-1}$ microturbulence value in a subset of 
models appropriate for lower gravity giants and supergiants. 
A comparison of model H$\gamma$ lines for the same 
$(T_{\rm eff}, \log g)$ parameters but with microturbulent 
velocities of 2 and 10 km~s$^{-1}$ shows only very small
differences in the profiles.  Thus, the hybrid treatment of
microturbulence in this application should have no 
significant impact on our results.

\subsection{Gravity Darkening} 

\citet{vonZeipel1924} found that the local energy radiated (and hence
the local temperature) of a rotating star varies as $T \propto
g^{1/4}_{\rm eff}$, so that the higher gravity pole is hotter than the
equator.   However, this can only be strictly correct in cases where the
gas is barotropic (pressure dependent on density only), which is
generally not the case for stars \citep{Rieutord2016}.  The full
solution to the problem requires consideration of the interior state and
motions (ESTER code; \citealt{EspinosaLara2013}). Fortunately
\citet{EspinosaLara2011} found an analytical representation of the
surface temperature variation with colatitude $\theta$ that matches
that from the detailed models quite well. We used this $\omega$-model in
our code by solving equations 18 and 23 in the development presented by
\citet{Rieutord2016} to determine the ratio of $T(\theta) / T_p$.  

Generally we adopt the $\omega$-model for our models, however, we
show an example of how the line profiles differ between the predictions
of the von Zeipel law and the $\omega$-model.  Figure~7 shows the
results for a model star with $R_p = 6.7 R_\odot$, $M = 20 M_\odot$,
$T_p = 40$ kK, $v_e = 598$ km~s$^{-1}$, and $v_c = 616$ km~s$^{-1}$,
i.e., a case close to critical rotation.  The images in the first column
show the surface brightness (specific intensity) in the far-ultraviolet
(1181 \AA ) where the contrast between the pole and equator is
especially striking.  The top image shows the limb darkened disk for the
corresponding non-rotating model, while the lower three images show the
same stellar model at near critical rotation described by the
$\omega$-model with inclination angles of $i=90^\circ$, $70^\circ$, and
$50^\circ$.  The next columns illustrate the appearance of several
surface-integrated flux profiles for these orientations and the two
gravity darkening laws considered.  The top row shows the non-rotating case for the
spectrum in the vicinity of \ion{C}{3} $\lambda 1175$, \ion{He}{1}
$\lambda 4471$, and \ion{He}{2} $\lambda 5411$. The next three rows give
the rotationally broadened model profiles plotted as a function of
Doppler shift relative to $v_{e} \sin i$.  If plotted versus actual Doppler
shift the profiles would appear narrower at lower inclination ($v_{e} \sin i
= 598$, 562, and 458 km~s$^{-1}$ for rows 2,3, and 4, respectively) due to
the inclination shift from equator-on, where the extreme rotation is
best observed, to pole-on, which shows no rotational broadening. 
However, by plotting the profiles relative to $v_{e} \sin i$, it is easier
to discern overall changes in the shape of the line profile with
inclination and between the predictions for the $\omega$-model (solid
lines) and von Zeipel law (dotted lines) for gravity darkening. 

\begin{figure*}
\includegraphics[angle=90,width=\textwidth]{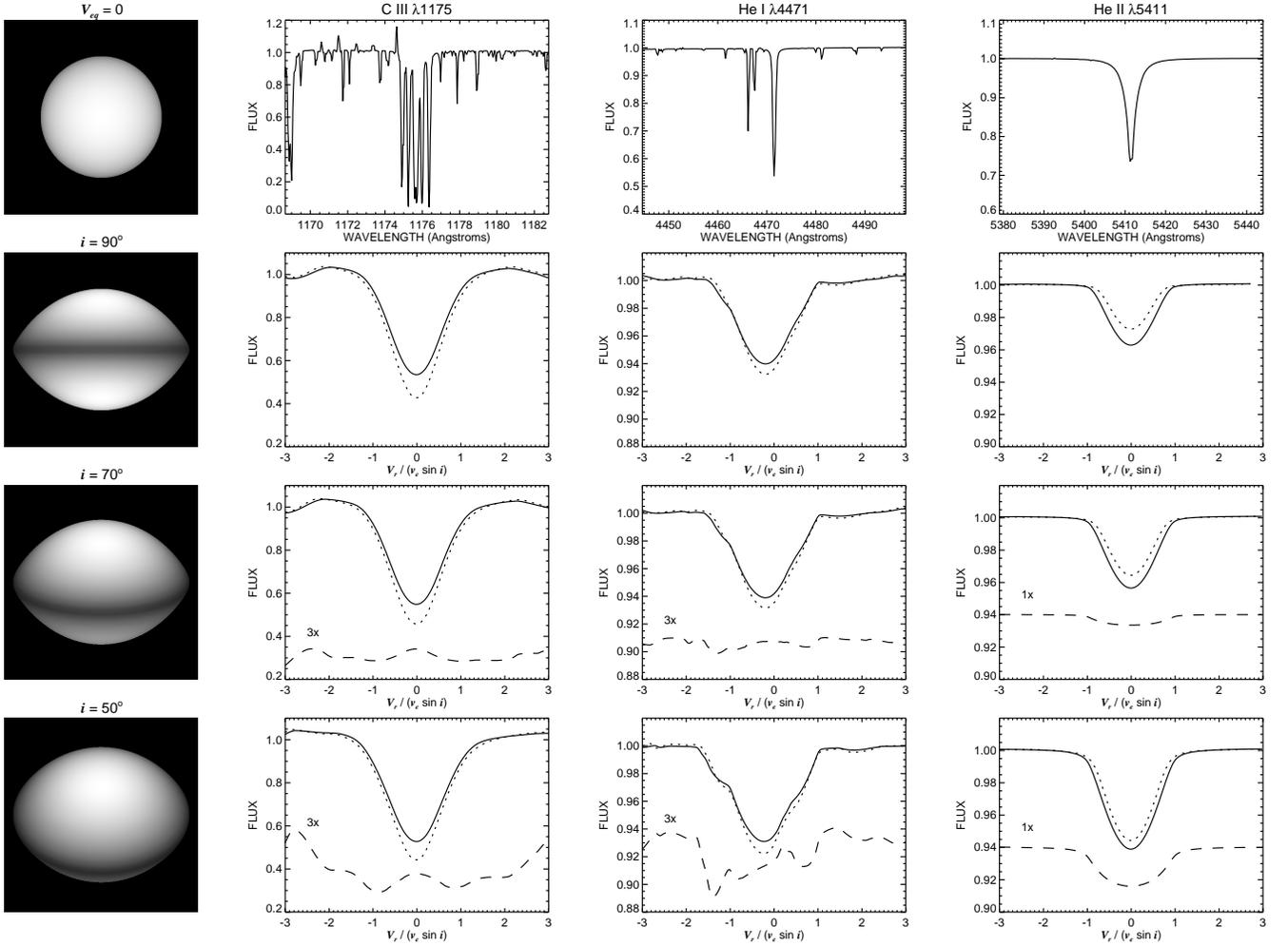}
\caption{A depiction of the appearance of a rapidly rotating star 
and a selection of its spectral lines as viewed at different inclinations.
The top row shows an image of a non-rotating star and its 
spectral lines in three regions.  The next three rows show a star at
near-critical rotation and the same three spectral features 
now plotted as a function of Doppler shift relative to $v_{e} \sin i$. 
Solid (dotted) lines show the model profiles using the $\omega$-model
(von Zeipel) descriptions of gravity darkening.  The dashed lines 
in the lower two rows show the difference between the $\omega$-model 
profiles for the specified inclination and that for $i=90^\circ$ (row 2).  
The difference is scaled up by a factor of three for the cases
of \ion{C}{3} $\lambda 1175$ and \ion{He}{1} $\lambda 4471$, indicated by a ``3x" label.}
\label{fig:gravity}
\end{figure*}

\placefigure{fig:gravity}

These models help to demonstrate how the differences between 
the gravity laws manifest in line profiles. 
The $\omega$-model predicts a pole to equator temperature variation 
that is less extreme than that from the von Zeipel prescription. 
This difference shows up best in the \ion{He}{2} $\lambda 5411$
profiles for $i=90^\circ$ (row 2, column 4).  In the von Zeipel 
case, the darker equator is more extensive and hotter pole more
limited in colatitude, and because the \ion{He}{2} line is 
preferentially formed in hotter environments, it appears weaker
in the von Zeipel case where hot conditions are confined to a 
smaller area.  The opposite is true of the other lines that grow
in strength at relatively cooler temperatures. 

A comparison of the relative changes in line shape and strength 
with inclination angle shows that some lines change significantly
(\ion{He}{2}) while others are approximately constant (\ion{C}{3}).
This demonstrates that an analysis of the rotational broadening 
among a sample of different line species can potentially help to 
constrain the value of the rotational inclination. 

\subsection{Transformation to the Observer's Frame} 

The derived model spectrum is created in units of physical flux 
in the rest wavelength frame of the star, and there are several 
steps required in order to compare the model directly with the 
observed line profile.  The first step is to shift the spectrum 
to the observed Doppler shift of the star. Initially we assumed 
a radial velocity of $+15$ km~s$^{-1}$ for $\zeta$~Oph 
\citep{Reid1993}, however we made a number of small revisions to
this estimate for the different lines in the sample. 
We adopted the mean of the radial velocities of VFTS~102 and
VFTS~285, but again we introduced small changes 
($< 20$ km~s$^{-1}$) on a line-by-line basis 
in order to align the model and observed line profiles. 
The model profiles were then transformed to the observed 
$\log \lambda$ wavelength grid by an integration scheme. 

The spectra were re-normalized to a unit continuum by 
selecting wavelength regions immediately to the blue and 
red of the main absorption feature, where we formed the 
ratio of observed-to-model fluxes in these regions.  
The model spectrum was then multiplied by a linear fit of the 
flux ratios in the rectification regions, so that the local 
continua of the observed and model spectra are in agreement.
The resulting model spectrum was then convolved with a 
Gaussian function to account for the minor amount of 
instrumental broadening associated with each spectrograph. 
The models of spectra for VFTS~102 were subject to a 
small reduction in line depth to account for the 
wavelength-dependent contribution of extra continuum flux
from the circumstellar disk (\S4.2).   
Finally the calculated portion of the model spectrum 
was inserted into an otherwise flat continuum spectrum 
outside of the wavelength range in the simulation, and 
these boundaries appear in some of the spectral plots below
where there appears to be a sudden jump to unity.  
  
\subsection{Parameter Fits} 

Our goal is to optimize the seven fitting parameters given in Table~4 in
order to best match the observed and model fluxes and rectified line
profiles. We found that we could converge to a unique solution through a
guided grid search method that relies primarily on the continuum fluxes to set
$E(B-V)$ and $R_p$, and then uses fits of the line profiles to help
determine $M$, $T_p$, $v_{e} \sin i$, and $i$. The final parameter is the He
abundance $y$ which we discuss separately in the next subsection.  

The procedure begins using assumed values for the parameters that are 
reasonably well known at the outset: $i$, $v_{e} \sin i$, $T_p$, and $y$. 
The first step is to compare the observed and model fluxes that span 
the full ultraviolet to optical range (including the circumstellar disk
flux contribution in the case of VFTS~102). We perform preliminary 
model simulations and check if there are any systematic trends in
the observed-to-model flux ratio as a function of wavelength.  If so,
then the reddening parameter $E(B-V)$ is revised in order to find
consistent observed-to-model flux ratios across the spectrum.  Next, we
consider the mean value of the observed-to-model flux ratio, and we
revise the polar radius $R_p$ in order to make this ratio unity for the
given values of distance and interstellar extinction (set through the
derived $E(B-V)$ and fixed $R_V$ values). 

The procedure next considers the fits of the spectral line profiles. 
A goodness-of-fit estimate is found by comparing the observed and model 
line profiles over a limited wavelength range that spans the full
absorption profile while excluding any problem regions that are marred
by background nebular or disk emission.
 This is particularly important in the case of VFTS 
102, where fits of the H and \ion{He}{1} lines were 
restricted to the extreme line wings to avoid emission 
components in the central parts of these profiles.
The scatter between the observed and model profiles
is compared to that in nearby continuum regions to find a reduced
chi-squared statistic $\chi^2_\nu$ for each spectral feature.  The code
also measures the ratio of (Observed - Calculated) / Observed equivalent
width to determine the sense of remaining discrepancies in the model,
and these are used to estimate the He abundance (\S4.7).  

The observed and model line comparison begins with the H-Balmer lines
that are sensitive to both temperature and gravity (through Stark
broadening) in the O- and B-type stars.  With the polar temperature and
radius set at this stage, the gravity dimension is explored through the
calculation of the Balmer lines for a test grid of model masses. 
For each test value of mass, we create model Balmer lines from an
integration of the surface with the local gravity set for 
each surface element based upon its radial distance from the center of the star and its rotational 
velocity, and then the corresponding specific intensity profile 
(with the associated Stark broadening) is derived from the 
pre-computed set (\S4.3).  Then we compare the model and observed
flux profiles to determine the goodness-of-fit for each Balmer line. 
A spline fit is made of the variation of the mean $\chi^2_\nu$ as a
function of assumed mass, and the minimum of this fit yields the
estimate of mass $M$.  

The basic procedure outlined above is repeated over a grid of test
values for polar temperature $T_p$, and the variation with assumed $T_p$
of the mean $\chi^2_\nu$ derived from all the lines in the sample is
used to find the best fit polar temperature.  The final step is to
conduct this complete analysis over a grid of assumed inclination $i$
and projected rotational velocity $v_{e} \sin i$, determine the
global minimum of the mean $\chi^2_\nu$, and find estimates 
for $i$ and $v_{e} \sin i$. 

We found that there were significant mismatches between the observed and
model profiles for certain line profiles that occurred in the analysis
of all three target stars.  These are systematic problems 
related to incomplete lists of possible line blends with other features,
non-standard abundances, and/or problems with the physical properties 
assigned to the atomic transitions in TLUSTY/SYNSPEC.  After some
experimentation, we limited the line sample to a set that gave mutually
consistent results, therefore any systematic errors that remain in the
analysis are treated consistently for all three stars.  The line set
adopted for the parameter estimation includes 
H$\zeta$ $\lambda 3889$, H$\gamma$ $\lambda 4340$, H$\beta$ $\lambda 4861$,
\ion{He}{1} $\lambda\lambda 3819, 4026, 4387$, and 
\ion{He}{2} $\lambda\lambda 4541, 4686, 5411$. 
We show in \S5 the fits for these nine features and for the
remaining seven excluded line profiles.  

\subsection{Helium Abundance} 

We noticed at the outset of the analysis that the He line model profiles 
calculated using specific intensity matrices based upon the solar He
abundance were often much weaker than the observed profiles. 
Consequently we computed additional model specific intensity matrices
using SYNSPEC for assumed He abundances of $2\times$, $3\times$, and
$4\times$ the adopted solar value.  The same parameter fitting procedure
was conducted for these different He abundances and the He line trends 
were examined in each case to determine how the models predicted
He line strengths that were systematically too weak and too strong. 

We show an example of the trends in Figure~8 for the case of
$\zeta$~Oph. The corresponding trends found for VFTS~102 and VFTS~285 are
qualitatively similar. Figure~8 shows the fractional differences in
(Observed - Calculated) / Observed line equivalent width as function of
assumed polar temperature $T_p$ for three \ion{He}{1} and three
\ion{He}{2} lines.  The \ion{He}{2} equivalent width ratios show a net
decline from underestimating the strength to overestimating the strength
with increasing $T_p$. The \ion{He}{1} $\lambda 4387$ equivalent width
ratio shows the opposite trend as expected, while the \ion{He}{1}
$\lambda\lambda 3819,4026$ ratios are approximately constant.  We
suspect that the latter two features are actually line blends that
change with temperature in differing ways so that the composite profile
is relatively constant (for example, the blend of \ion{He}{1} $\lambda
4026$ and \ion{He}{2} $\lambda 4025$). The large plus sign near the
center marks the average position of all the \ion{He}{1} and \ion{He}{2}
trend crossings.  This occurs at $< W_\lambda (O-C)/O > = +0.069 $ for
the $2\times$ solar model (left panel; He too weak) and at $< W_\lambda
(O-C)/O > = -0.062 $ for the $3\times$ solar model (right panel; He too
strong).  Thus, the best match of the He line strengths occurs for an
intermediate He abundance between these cases.

\begin{figure*}
\plottwo{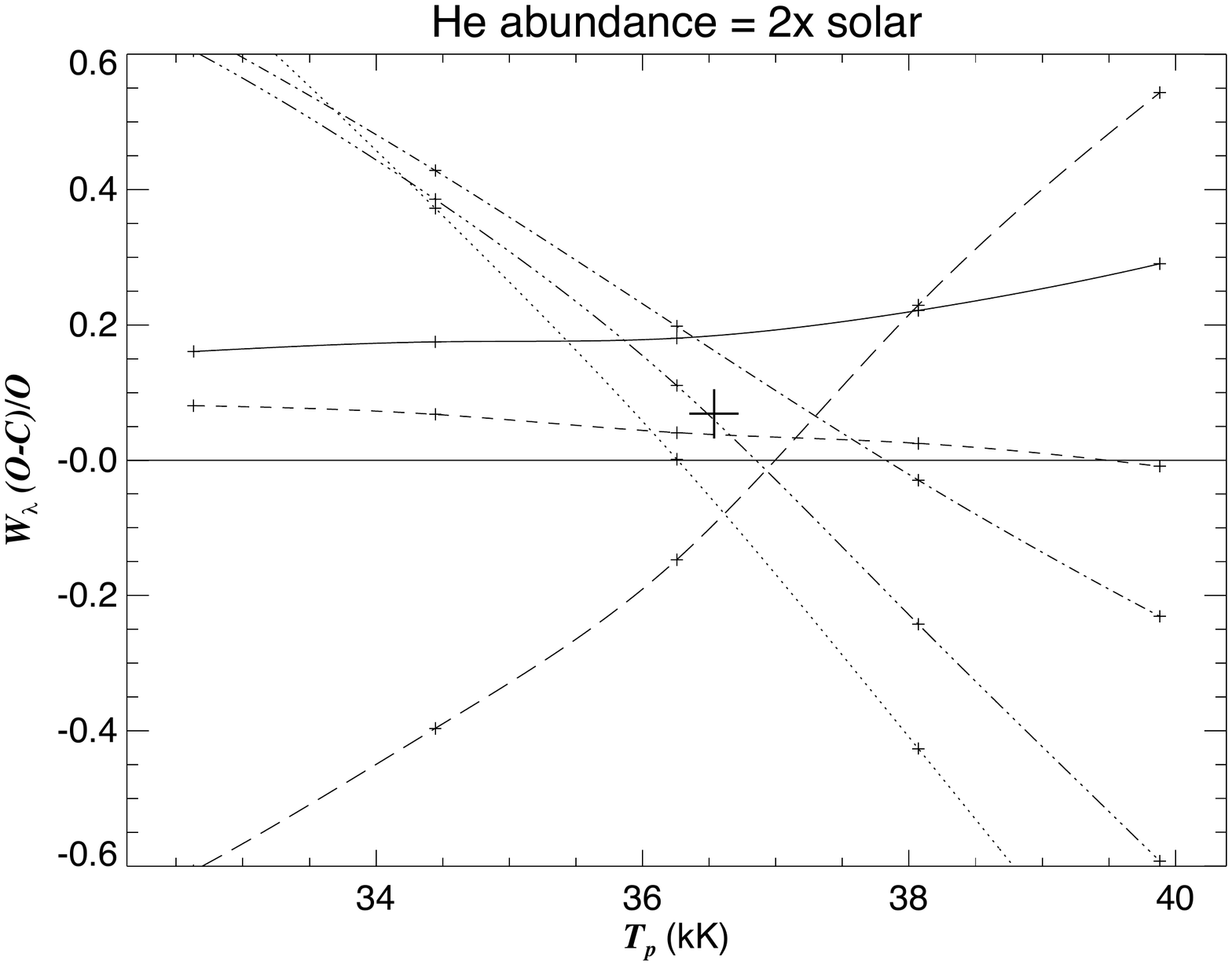}{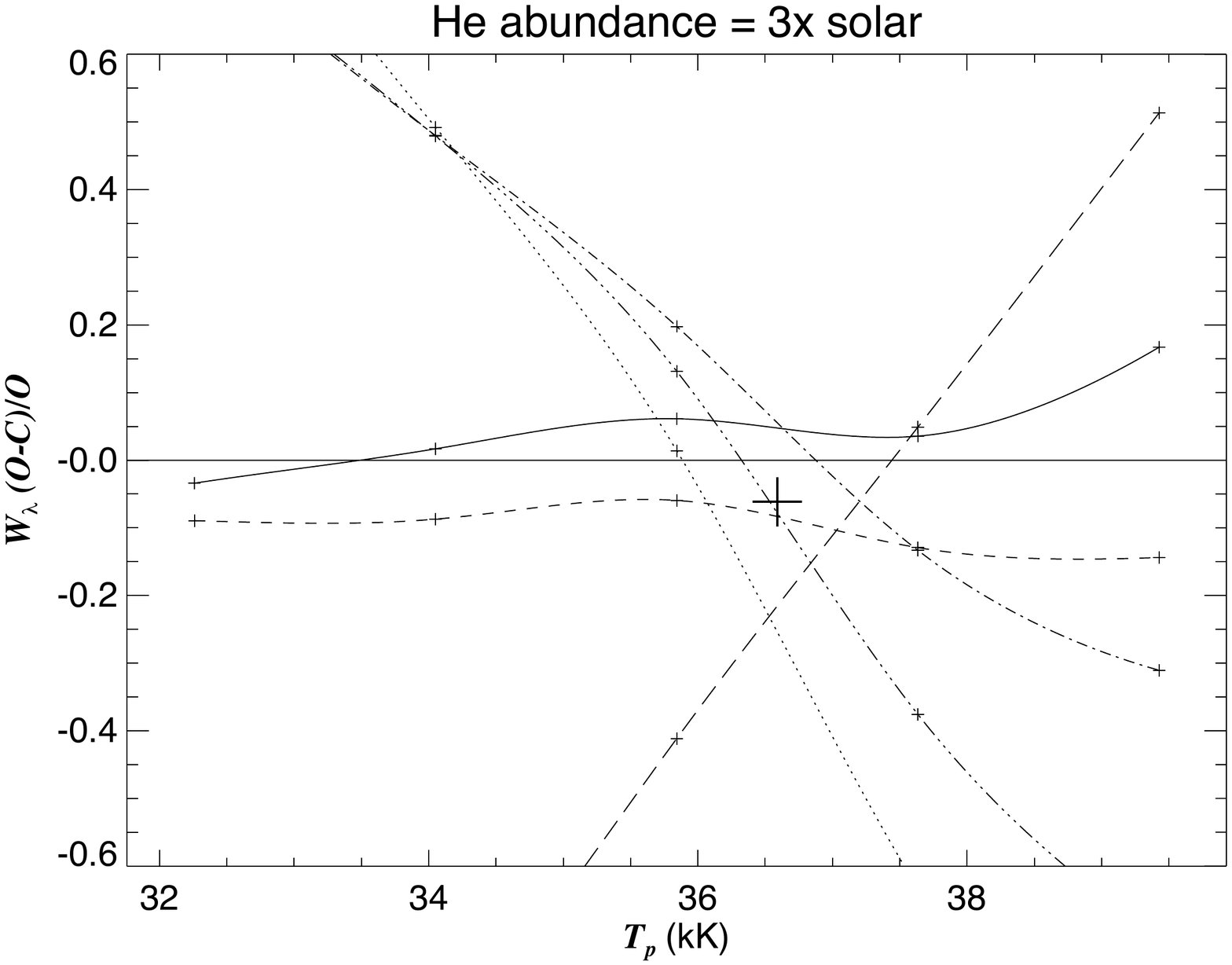}
\caption{
{\it Left:} A plot of the fractional differences between the observed 
and model calculated equivalent widths for a He abundance that is
twice solar.  The model results for each test polar temperature $T_p$ 
are plotted as small plus signs that are connected by spline fits. 
The \ion{He}{1} $\lambda\lambda 3819, 4026, 4387$ features are plotted as 
solid, dashed, and long dashed lines, respectively, and the \ion{He}{2} 
$\lambda\lambda 4541, 4686, 5411$ features are plotted as dotted, dot-dashed, 
and triple dot-dashed lines, respectively.  The large plus sign near 
the center marks the mean of the \ion{He}{1} and \ion{He}{2} 
intersection points, which is positive in this case (He too weak). 
{\it Right:} The same trends plotted for an assumed He abundance 
that is three times the solar value that leads to a negative mean 
(He too strong).}
\label{fig:helium}
\end{figure*}

\placefigure{fig:helium}

We made similar plots for all four test cases of He abundance $y$,
and we used a spline fit of $(y, < W_\lambda (O-C)/O >$ to find 
the zero-crossing position that corresponds to the best fit of 
the helium abundance based upon these six He lines.  
The results are listed with the other parameters 
in Table~4 with uncertainties based upon the scatter in the 
\ion{He}{1} and \ion{He}{2} intersection points in these plots. 

\subsection{Parameter Uncertainties} 

The predominant source of uncertainty in the parameter estimations comes
from the spread in derived parameter values for fits of the individual 
lines in the default set of three H, three \ion{He}{1}, and three
\ion{He}{2} features.  These are systematic errors related to the model
itself, so in most cases we have estimated the parameter uncertainties
from the line-to-line standard deviation of the results found using the
individual line fits.  There are two other sources of significant
uncertainty that must also be considered.  The fractional uncertainty in
distance is largest for the case of $\zeta$~Oph ($12\%$), therefore this
is the most important factor in assessing the uncertainty in polar
radius $R_P$ which is linearly dependent on the assumed distance. The
other key element in the uncertainty is the interstellar extinction
that depends on the assumed value of the ratio of total-to-selective
extinction $R_V$.  The reddening $E(B-V)$ is modest for $\zeta$~Oph and
VFTS~285, so the underlying uncertainty in $R_V$ has only a small affect
on the results.  However, the reddening of VFTS~102 is much larger, so
uncertainties in $R_V$ are important. We tested the sensitivity of the
results by making simple SED fits (like those in Fig.~4) 
for both the adopted value of 
$R_V=2.76$ and the nominal value of $R_V=3.1$, and we found that the
derived angular diameter increased by $28\%$ using the latter value. 
The polar radius $R_p$ varies directly with angular size, so we included
this factor in the final uncertainty estimate for $R_p$ for VFTS~102.  

The bottom two rows of Table~4 list the statistics associated with the fits. The reduced chi-square $\chi^2_\nu$ is the average of the
individual chi-square measurements 
for all the lines used in the sample.  It can be somewhat misleading
given the complicated and often deceiving nature of extreme rotational
broadening. For example, $\chi^2_\nu$ is the smallest for fits of the
lines of VFTS~102, but this is mainly the result of extremely shallow
lines that have a depth not much larger than the scatter in the
continuum. The final row reports the standard deviation of the observed
to model flux ratio for all 16 spectral regions from the FUV to the optical,
and it provides a sense of the success of the flux fits (worse in the
case of VFTS~102 where complications exist due to the flux of the
circumstellar disk).  The fits associated with the parameter estimates
in Table~4 are discussed in the next section. 

\section{Spectrum Synthesis Fitting Results}  

\subsection{$\zeta$ Oph}  

\begin{figure*}
\includegraphics[width=\textwidth]{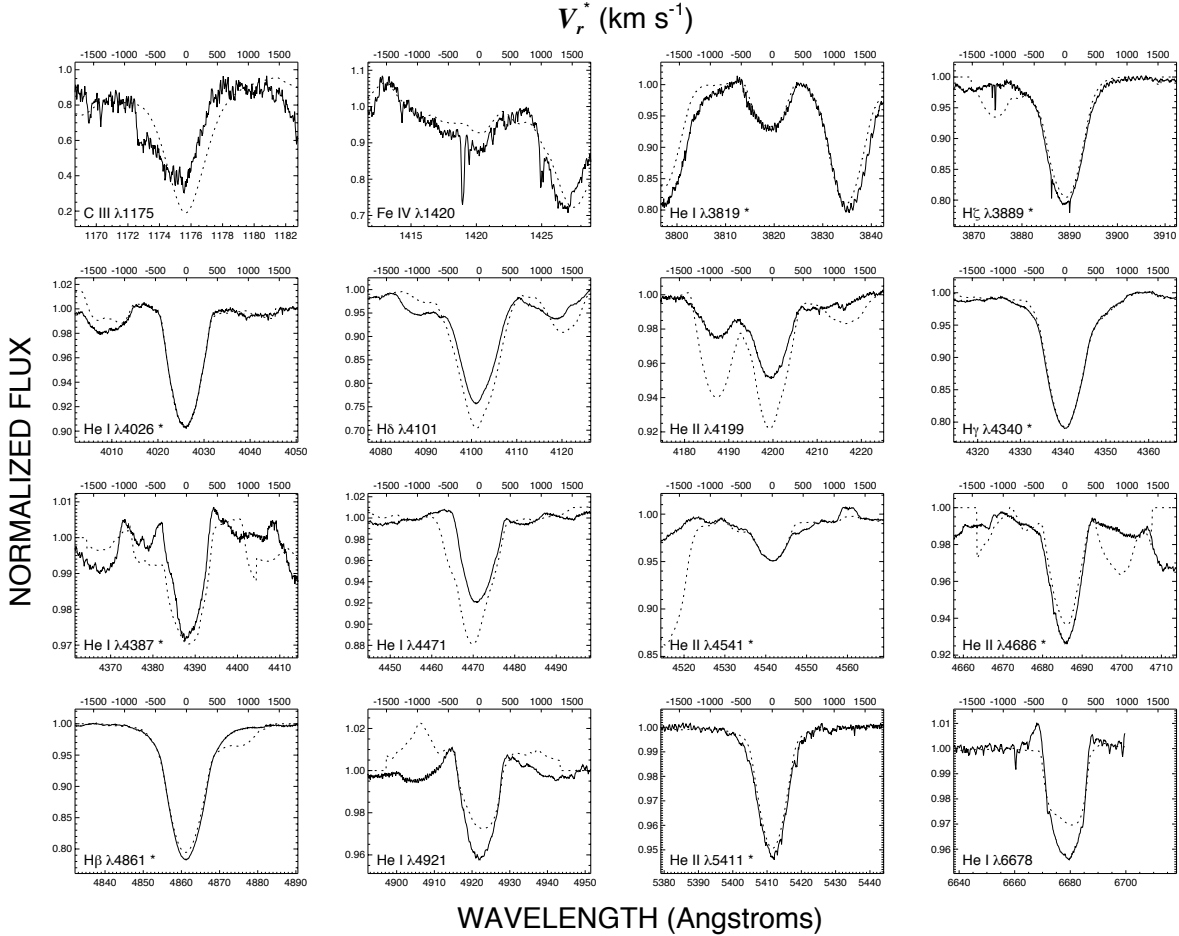}
\caption{
The continuum normalized spectral features of $\zeta$~Oph (solid lines) 
together with the synthetic spectra from the rotational model (dotted lines). 
The top axis represents Doppler shift in the rest frame of the star, 
while the lower axis depicts the observed (heliocentric) wavelength. 
The primary feature in each panel is identified with a label in 
the lower left, and those features included in the parameter fitting 
scheme are indicated by an asterisk appended to the label. 
}
\label{fig:speczo}
\end{figure*}

Figure \ref{fig:speczo} shows the 16 spectral features that we modeled
with the rotation code described in \S4.  The default set of nine lines
used in the parameter fitting code (indicated by asterisks in the
identifying labels) are generally well fit by the model, however
discrepancies from the fit in other cases deserve some comment.  The
\ion{C}{3} $\lambda 1175$ feature appears to show a significant
blue-shift compared to the model, and we suspect that this transition is
partially influenced by the stellar wind, appearing like a weak P~Cygni
feature (observed as a wind feature in the O-star binary UW~CMa;
\citealt{Drechsel1981}).  The other ultraviolet feature is the
\ion{Fe}{4} $\lambda 1420$ blend that appears to be slightly too weak in
the model, perhaps due to the choice of microturbulence in the TLUSTY
model or to uncertainties in the atomic oscillator strengths.  The
H$\delta$ $\lambda 4101$, \ion{He}{1} $\lambda 4471$, and \ion{He}{2}
$\lambda 4199$ lines all appear to be consistently too deep in the model
(including the \ion{C}{3} $\lambda 4186$ line in the blue wing of the
latter), so they were excluded from the parameter fit. The longer
wavelength transitions of \ion{He}{1} $\lambda\lambda 4921, 6678$ appear
to show disk-like emission in their extreme wings (possibly also present
in \ion{He}{1} $\lambda 4387$), so they were also omitted from the fit.
The spectrum of $\zeta$ Oph does occasionally exhibit double-peaked
emission like that of the disk emission observed in Be stars \citep{Vogt1983},
but no H$\alpha$ emission is apparent in the $\zeta$~Oph spectra used
here.   

\placefigure{fig:speczo}

Our motivation for including $\zeta$ Oph in this study was to test our
parameter fitting results with those obtained independently using a
similar code by \citet{Howarth2001}.  The two sets of results are
compared in columns 2 and 3 of Table~4. There is good
agreement in most of the derived parameters.  The polar temperature
$T_p$ derived by \citet{Howarth2001} is slightly higher due to our use
of fully line-blanketed model atmospheres that tend to assign lower
$T_{\rm eff}$ than models based on H and He line opacities alone
\citep{Lanz2003}.  The other difference is the use of the von Zeipel law
for gravity darkening used by \citet{Howarth2001}. 
The equator to polar temperature contrast is larger in the von Zeipel description versus the $\omega$-model for a given rotation rate, and we arrive at the same ratio of $T_e/T_p$ as derived by \citet{Howarth2001} (who utilized the Von Zeipel law) by using a relatively larger angular velocity ratio ${\Omega} / {\Omega}_c$ when using the $\omega$-model.

Our derived He abundance $y=0.24 \pm 0.07$ is the same within
uncertainties as found by \citet{Howarth2001}, $y=0.20 \pm 0.03$,
confirming the apparent He enrichment in the atmosphere of $\zeta$~Oph.
Models that neglect the changes of the star's 
shape and gravity darkening due to rotation tend to
arrive at a lower He abundance: for example, $y=0.16$ from
\citet{Herrero1992}, $y=0.11$ from \citet{Villamariz2005}, and $y=0.10 -0.12$ from \citet{Cazorla2017}.   The fast projected rotational
velocity we find is similar to that found in most other studies, with
the exception of \citet{Simon-Diaz2014}, who split the apparent
broadening between rotation, $v_{e} \sin i = 303 - 319$ km~s$^{-1}$, and
macroturbulence, $v_{\rm m} = 159$ km~s$^{-1}$.

\subsection{VFTS 102}  

\begin{figure*}
\includegraphics[width=\textwidth]{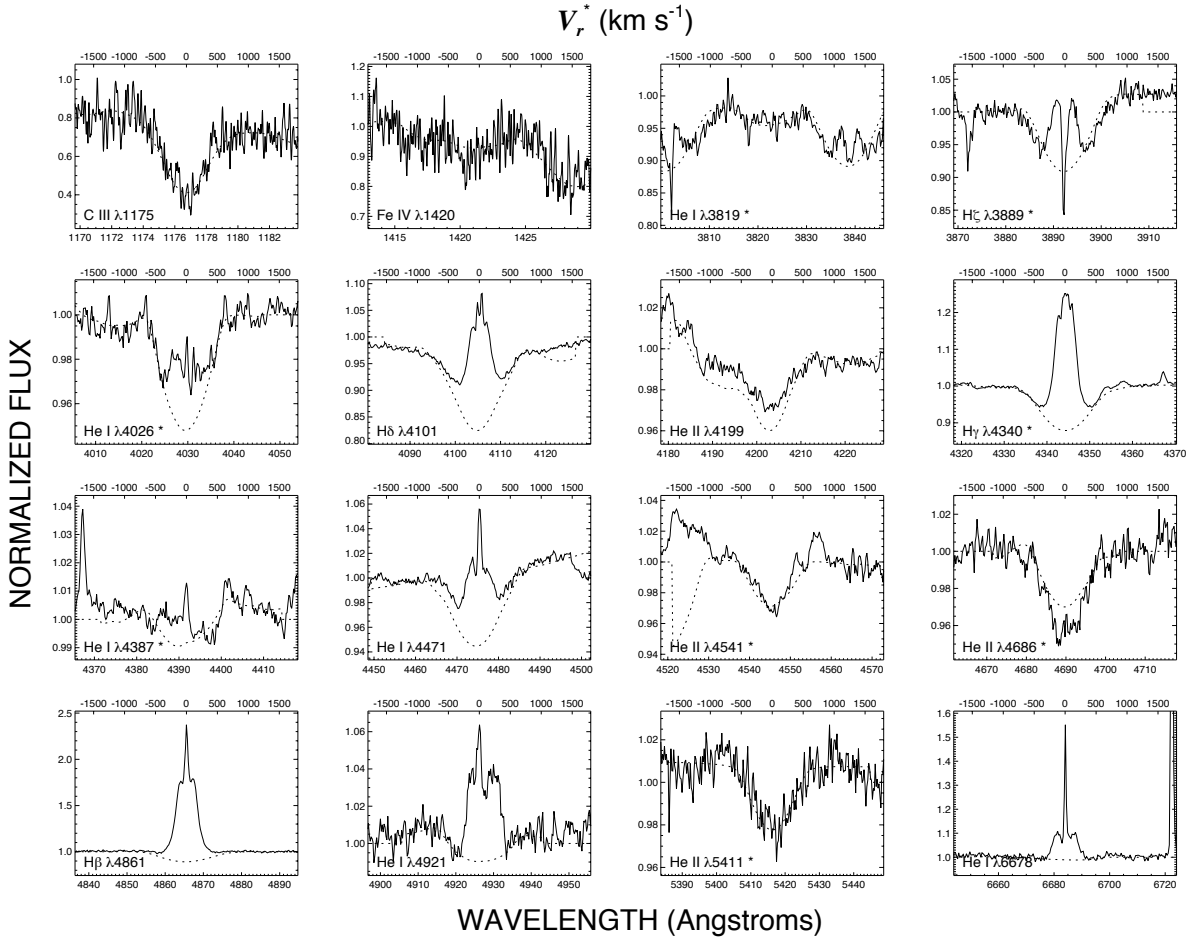}
\caption{
The continuum normalized spectral features of VFTS~102 (solid lines) 
together with the synthetic spectra from the rotational model (dotted lines) in the same format as Figure \ref{fig:speczo}. The sharp features in the cores of some of the H Balmer and \ion{He}{1} lines are artifacts from incomplete removal of the surrounding nebular emission lines. 
}
\label{fig:spec102}
\end{figure*}

Many of the spectral line features of VFTS~102 are altered in some way
by the presence of a well-developed circumstellar disk.  We discussed
above in \S4.2 how disk emission adds to the spectral energy
distribution at longer wavelengths and how the disk continuum acts to
make the spectral lines appear shallower.  The other striking aspect is
how the disk emission appears as a new profile component in the core of
the H Balmer and \ion{He}{1} lines as shown in Figure \ref{fig:spec102}.
For these cases, the observed and model profile goodness-of-fit
statistic $\chi^2_\nu$ was calculated only for the line wing portions of
the profile where no disk emission appears.  This approach is 
successful in showing the presence of residual disk emission that is not
obvious on first inspection (for example, the \ion{He}{1} $\lambda 4026$
profile in Fig.\ \ref{fig:spec102}).  Making a fit of the line wings
failed in the case of H$\beta$ $\lambda 4861$ because disk emission
extends into the far wings, so this feature was omitted from the
parameter fitting procedure.  This leaves only fits of H$\zeta$ $\lambda
3889$ and H$\gamma$ $\lambda 4340$ for the determination of the mass
through the apparent Stark (collisional) broadening.  Tests showed that
omission of H$\beta$ in fits of the spectra of $\zeta$~Oph and VFTS~285
changed the derived mass by less than $1\%$, therefore we doubt the
omission of H$\beta$ has any significant impact on the final parameter
solution for VFTS~102. 

\placefigure{fig:spec102}

The derived average parameters in Table~4 all agree 
within uncertainties with those estimated in the discovery paper by 
\citet{Dufton2011} (see their Table~1).  In particular, the estimate 
of projected rotational velocity $v_{e} \sin i = 649 \pm 52$ km~s$^{-1}$
is the largest among all three stars, and the equatorial velocity is 
the same as the critical velocity within errors.   Thus, VFTS~102 
appears to be a star that has attained critical rotation. 
This extraordinary spin probably assists those mass loss processes 
that feed gas into the circumstellar disk, creating a vigorous 
decretion disk despite the disk gas ablation that occurs due to 
the harsh radiation field of the star \citep{Kee2016}. 

\subsection{VFTS 285} 

The rotational model parameter fits for VFTS~285 indicate that the 
star is the hottest and most massive of the three targets.  
The star's true equatorial velocity is about 
the same as that of VFTS~102, however 
because the star is more massive, the critical velocity is higher, 
and therefore the star has a sub-critical spin, ${\Omega} / {\Omega}_c =
0.95$. The hotter temperature and slower spin relative to critical 
rotation are probably the reasons why no circumstellar disk is 
found for VFTS~285 (in contrast to the case of VFTS~102; \S5.2).  
The average parameters given in column 5 of Table~4
agree within errors with those derived from \citet{Sabin2017}
that do not include rotational deformation in the model.  
The one exception is the He abundance
that \citet{Sabin2017} find to be only somewhat enhanced, 
$y=0.14$ compared to our result of $y=0.34 \pm 0.14$ (the largest
He overabundance among the three stars).  This same kind of difference 
was noted above between rotating and non-rotating physical model results 
for $\zeta$~Oph (\S5.1). 
In the non-rotating models \ion{He}{2} line formation occurs 
over the entire visible hemisphere, while in the rotating 
models that include gravity darkening, \ion{He}{2} line
formation is more restricted to the hotter polar zones 
(because the \ion{He}{2} lines weaken in the cooler 
equatorial zone).  Consequently, in order to match the 
observed line strength, the rotating models compensate for
the smaller area of formation by increasing the He abundance. 
The other difference in our work is the neglect of stellar 
winds in the TLUSTY models. The \ion{He}{2} $\lambda 4686$ line
that we use is sensitive to wind emission in more luminous stars
\citep{Walborn1971}, but in the case of VFTS~285, the model
\ion{He}{2} $\lambda 4686$ line appears to match the 
observed line as well as found for the other \ion{He}{2} lines
(Fig.\ 11).
 
\begin{figure*}
\includegraphics[width=\textwidth]{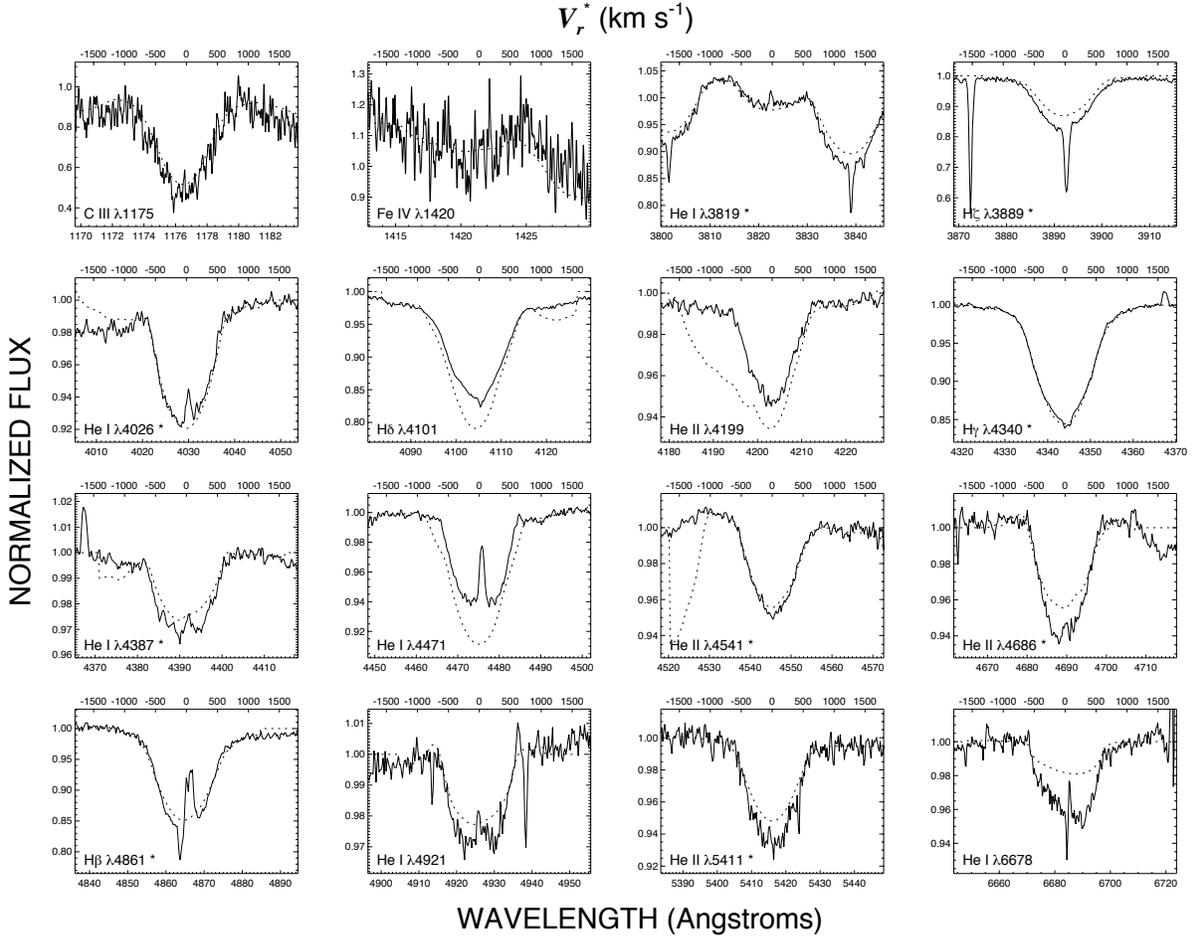}
\caption{
The continuum normalized spectral features of VFTS~285 (solid lines) 
together with the synthetic spectra from the rotational model (dotted
lines) in the same format as Figure \ref{fig:speczo}. The sharp features
in the cores of some of the H Balmer and \ion{He}{1} lines are artifacts
from incomplete removal of the surrounding nebular emission lines. 
}
\label{fig:spec285}
\end{figure*}

The spectral line fits shown in Figure \ref{fig:spec285} are 
mostly satisfactory among the default set (marked by asterisks in 
the figure panels), except in the cores of some of the H Balmer 
and \ion{He}{1} lines where sharp features remain from over- or 
under-subtraction of the nebular emission from the surrounding gas. 
These core regions were excluded from the goodness-of-fit measurements. 

\placefigure{fig:spec285}

We decided to experiment with model fits of the spectral features of 
VFTS~285 by changing the gravity darkening prescription to 
the von Zeipel law in order to demonstrate how the choice of 
gravity darkening influences the solution.  The resulting parameters
using the von Zeipel law are shown in the final column of Table~4. We found that the model predicted \ion{He}{1} and
\ion{He}{2} line profiles that were still too weak ($7\%$) compared to
the observed profiles even with the largest assumed He overabundance, 
$y=0.4$ (\S4.7). Thus, applying a rotational model using the von
Zeipel law appears to lead to an over-estimate of He abundance in this
case.  Rather than extrapolate to even higher He abundances, we simply
report the results of the $y=0.4$ fits in Table~4.  
The von Zeipel model is best fit with a star that is rotating somewhat closer to the critical rate with greater equatorial extension and a larger range in the polar to equatorial temperature. 

\section{Evolutionary Origins} 

\subsection{Single Star Models} 

Both VFTS~102 and VFTS~285 display exceptionally large rotational 
line broadening compared to other O-type stars in the VFTS sample 
\citep{Ramirez2013}.  Here we consider what processes may have 
contributed to their extreme spins.  The first possibility is 
that both are very young stars that attained their 
rapid spin due to accretion of their natal disks.  
\citet{Ekstrom2008} and \citet{Brott2011} calculated evolutionary 
tracks for stars born as rapid rotators, and in Figure~12 we 
show evolutionary tracks in the Hertzsprung-Russell Diagram (HRD) 
for three massive stars from \citet{Brott2011}.   These particular 
tracks were made assuming LMC abundances and initial equatorial 
velocities of $v_e \approx 550$ km~s$^{-1}$.  The track for the 
$16 M_\odot$ model shows the normal evolution to higher luminosity
and cooler temperature as core H-burning concludes, but the tracks 
for $19 M_\odot$ and $25 M_\odot$ show evolution to higher temperatures. 
This behavior in massive, fast rotators is due to extensive mixing in the interior 
that replenishes the H core and dredges up the processed He into the envelope.
Thus, mixing tends to homogenize the composition of the core and envelope.

\placefigure{fig:CCFs of FUV spectra}
\begin{figure*}
\includegraphics[angle=90,width=\textwidth]{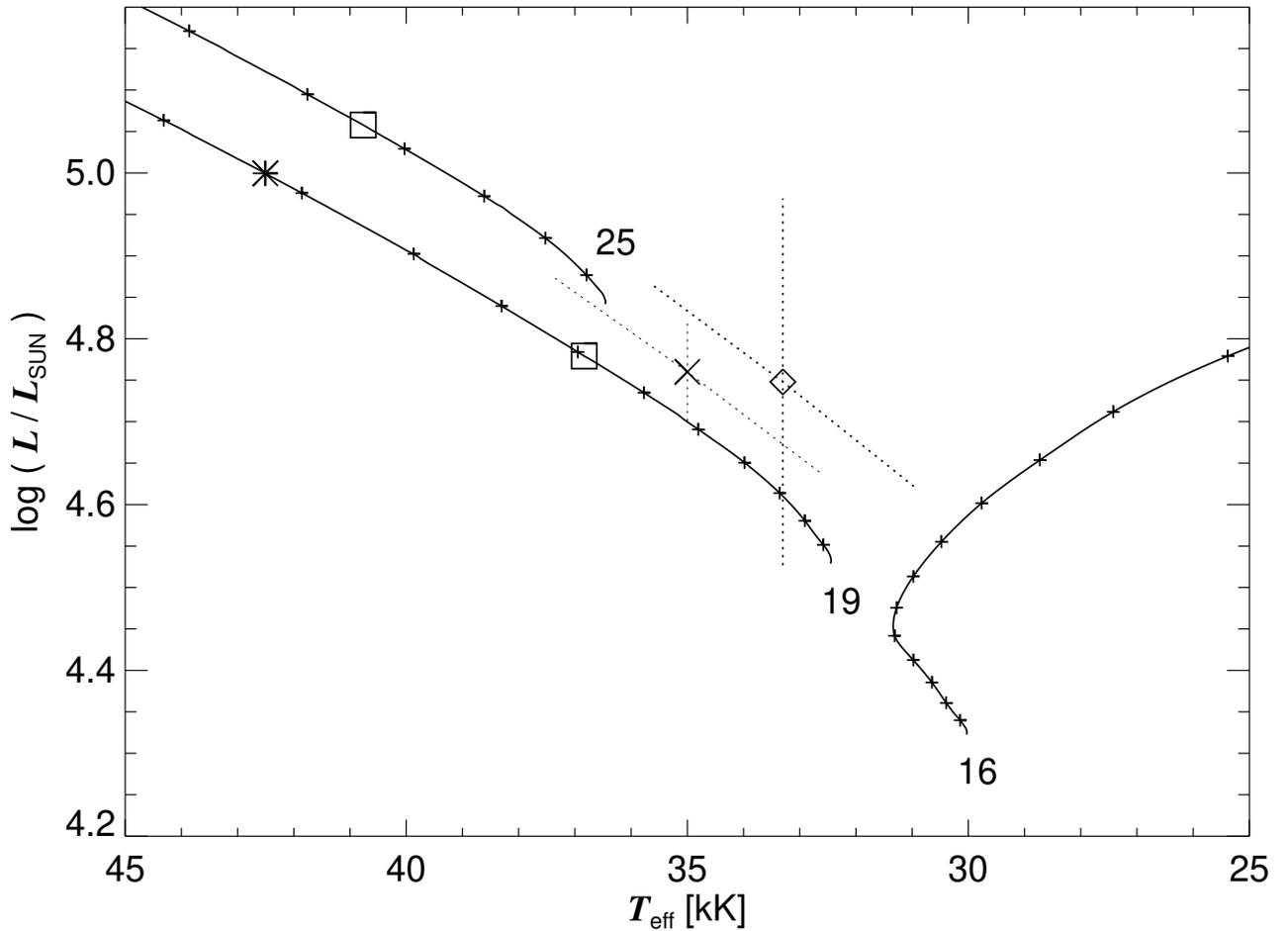}
\caption{Evolutionary tracks in the HRD for rapidly rotating massive stars
from \citet{Brott2011}.  The solid lines show the tracks for 
stars of masses $16 M_\odot$, $19 M_\odot$, $25 M_\odot$ with 
initial equatorial velocities of 562, 557, and 548 km~s$^{-1}$,
respectively.  Small plus signs indicate time intervals of 1 Myr,
and the square and asterisk symbols show the point on the tracks
where the surface He abundance reaches $y=0.2$ and 0.4, respectively. 
The diamond and X symbols mark the observed average temperature 
and luminosity for VFTS~102 and VFTS~285, respectively, from 
model fits of the spectral lines. 
}
\label{fig:brott}
\end{figure*}

We also plot in Figure~12 the derived estimates of $<T>$(all) and $\log L/L_\odot$ from Table~4 for VFTS~102 and VFTS~285.  Note that the $T_{\rm eff}$ estimates for the models of \citet{Brott2011} are based upon an average over the surface of the star assuming a von Zeipel gravity darkened flux, so they are not exactly comparable to our results for the $\omega$-model (but the difference is small; see the two gravity darkening cases for VFTS~285 in Table~4). Furthermore, the model equatorial velocities are somewhat smaller than our estimates for the stars, 
so trends related to rotation might be even more extreme for evolutionary tracks at higher rotation speeds. 
The position of VFTS~102 is somewhat overluminous for its estimated mass ($18 \pm 6 M_\odot$) 
and is slightly cooler than predicted for the age of 
the nearby stars in the vicinity of the LH~99 OB association 
of 5.7 Myr \citep{Schneider2018}.
Furthermore, an enhanced He abundance of $y=0.2$ is obtained at an older age of 6.9 Myr (or longer for masses lower than $19 M_\odot$) and at a hotter temperature than is observed for VFTS~102.  
The position of VFTS~285 is cooler and less luminous than predicted for its estimated mass ($28 \pm 8 M_\odot$), and the observed He over-abundance of $y=0.34 \pm 0.14$ only occurs at much higher temperatures in the $25 M_\odot$ track. This comparison between the observed and model tracks for VFTS~102 and VFTS~285 indicates that these stars are likely different than predicted for models of stars with fast rotation at birth. 

Stars may also become rapid rotators towards the terminal age main 
sequence (TAMS) stage due to the transfer of angular momentum from 
core to surface by meridional currents \citep{Ekstrom2008}.   
However, this spin-up occurs over a relatively short time near the TAMS, 
and both VFTS~102 and VFTS~285 appear to be too young to have 
reached the TAMS.  Consequently, we doubt that their fast rotation
is related to mixing and spin-up associated with the TAMS phase. 

\subsection{Binary Star Models} 

There are several processes involving interacting binary stars 
that can lead to the spin up of the mass gainer star \citep{deMink2013}.
Binary systems born with short periods will probably enter 
a mass transfer stage during slow core-H burning expansion (Case A), 
and in many circumstances this will result in a merger through 
a common envelope event (CEE).   Population synthesis models by 
\citet{deMink2014} suggest that $8\%$ of massive stars drawn from 
a constant star formation sample are, in fact, such merger products.
\citet{Menon2021} present models for binary mergers in the LMC,
and they show that many binaries with initial periods less than 
two days will produce a merger. Our understanding of the physical
processes leading up to a CEE is still developing \citep{Ivanova2013}, however the
merged star is likely to exhibit rapid rotation, equatorial mass loss,
an enriched surface He abundance, and overluminosity for its mass
\citep{Ivanova2003}.  The properties of the merger product depend
critically on the extent of He dredged up into the envelope and the time
since the merger \citep{Glebbeek2013}. One key example is the hot merger remnant
model that is successful in describing the properties of the B-supergiant
progenitor of SN 1987A \citep{Podsiadlowski1992,Menon2017}. 
Another example is the magnetic star $\tau$ Sco that may have formed 
through a merger that generated a strong magnetic field as described by
\citet{Schneider2019,Schneider2020}. Their models suggest that
the merger product will spin-down on a thermal timescale ($\sim 10^4$ years) 
as a result of redistribution of angular momentum in the stellar interior. 

Massive binaries with periods greater than a few days will begin mass
transfer as the larger mass donor star expands in the H-shell burning
stage (Case B).  Mass transfer will lead to the spin up of the mass
gainer which becomes a rejuvenated star of larger mass than it initially
started with \citep{deMink2014}. When the donor explodes as a supernova,
the binary may become unbound (if the donor mass remained large after
the mass transfer episode or the donor experiences an asymmetric kick
during the SN) or the SN remnant may remain to orbit the mass gainer.
The latter systems are observed as massive X-ray binaries in which 
reverse mass transfer occurs to the neutron star or black hole
remnant. Both circumstances will impart a runaway velocity to the
surviving gainer that is comparable to the orbital velocity at the time
of the SN. Many of the fast-moving OB runaway stars appear to be rapid
rotators and are often single stars
\citep{Blaauw1961,Blaauw1993,Gies1986,Hoogerwerf2001,Platais2018,Schneider2018}. 
     
\subsection{Dynamical Processes} 

The central region of the R136 cluster in the 30 Dor region has a very 
high spatial number density of massive stars
\citep{Massey1998,Crowther2019}, and dynamical encounters between stars
and binaries can play an important role in their evolution.  In rare
cases, a physical stellar collision can lead to the formation of a
rapidly rotating star with properties similar to those formed by a close
binary merger \citep{Sills2005,Fujii2012}.  Gravitational encounters
between wider binary and single stars (and binary and binary stars)
offer another way to eject a high velocity star through an interaction
that transforms the orbital binding energy of a target binary into the
kinetic energy of the escapee \citep{Gualandris2004}. 

In the following subsections, we will compare the predictions from these
different processes with the observed properties of the three rapid 
rotators investigated in this paper. 

\subsection{$\zeta$ Oph} 

$\zeta$~Oph is the closest Galactic O-star and it is a well-known 
runaway star \citep{Blaauw1961}.  Its trajectory across the sky shows 
that it was ejected from the Upper-Centaurus-Lupus (UCL) Association \citep{Hoogerwerf2001}.
\citet{vanRensbergen1996} argue that it was a member of an interacting 
binary system that spun up the mass gainer ($\zeta$~Oph) prior to a 
SN explosion that disrupted the system and imparted a runaway velocity. 
The fact that there is no evidence of orbital motion \citep{Gies1986}
is consistent with its status as a single star.  \citet{Neuhauser2020}
presented an analysis of the motions of $\zeta$~Oph and nearby pulsars, 
and they argue that the SN that created the radio pulsar PSR~B1706-16
caused the ejection of $\zeta$~Oph, in addition to 
the release of a significant amount
of $^{60}$Fe gas (some of which was eventually captured on Earth).  Thus,
$\zeta$~Oph is a prime example of a star that was spun up to near
critical rotation (${\Omega} / {\Omega}_c = 0.95$) by mass transfer from
a companion that exploded as a SN and imparted a runaway velocity to the
survivor.

\subsection{VFTS 102} 
        
The SN ejection mechanism that explains the properties of $\zeta$~Oph 
was explored as the origin of the rapid rotation of VFTS~102 in 
the discovery paper by \citet{Dufton2011}.  They noted that
the nearby pulsar PSR~J0537-6910 is surrounded by an X-ray 
emitting bow shock that appears to be directed away from the position 
of VFTS~102, and this implies that the pulsar is a runaway object from 
the vicinity of VFTS~102.  Furthermore, they argued that their 
measurement of radial velocity, $228$ km~s$^{-1}$, was sufficiently 
different from the mean for the region that VFTS~102 was also a 
runaway object.  However, the runaway status of VFTS~102 is 
controversial.   
Our derived average radial velocity 
is $267 \pm 3$ km~s$^{-1}$, and this is the 
same within errors as the mean for stars 
in the region around the LH~99 association, 
$274 \pm 13$ km~s$^{-1}$ \citep{Evans2015}.   
Furthermore, the apparent proper motions of VFTS~102 
from Gaia EDR3 \citep{Gaia2021} are 
$\mu_{RA} = 1.73 \pm 0.04$ mas~yr$^{-1}$ and 
$\mu_{DEC} = 0.71 \pm 0.03$ mas~yr$^{-1}$ which agree 
with the mean values for other nearby massive stars. 
For example, we formed a sample of 65 O-type stars 
within a 2 arcmin separation from the massive star Brey~73,
which lies near the center of the OB-association LH~99 
that is close in the sky to VFTS~102 \citep{Lortet1991}.
The mean proper motions of these stars from Gaia EDR3 are 
$\mu_{RA} = 1.63 \pm 0.16$ mas~yr$^{-1}$ and 
$\mu_{DEC} = 0.63 \pm 0.18$ mas~yr$^{-1}$, i.e.,
the same within errors as those for VFTS~102.
These results suggest that VFTS~102 is not a
runaway star. The difference between its spatial velocity 
and that of the LH~99 association stars is no more than 
$\approx 30$ km~s$^{-1}$, which is smaller than we 
would expect from a runaway star.

\citet{Jiang2013} presented a binary merger model for VFTS~102, and they
showed how a close binary with an initial primary star mass of $12 - 15
M_\odot$, mass ratio $M_2/M_1 > 0.63$, and orbital period $P < 1.5$ d
can evolve into contact and merge to create a rapidly rotating star.  
These parameters are consistent with the current mass of $18 M_\odot$ 
provided some mass loss occurred during the CEE. 
VFTS~102 does display the properties predicted for a merger: 
very rapid rotation, enhanced He abundance, overluminosity, and 
evidence of ongoing mass loss into a large equatorial gas disk. 
Furthermore, our analysis suggests that it is rotating at essentially 
the critical rate, so little time has elapsed since the spin-up event
for active spin-down processes to occur that are related to evolution 
\citep{Brott2011}, wind mass loss \citep{Gagnier2019}, angular 
momentum loss into the circumstellar disk \citep{Krticka2011},
and internal restructuring \citep{Schneider2020}. 
These facts, combined with the lack of a substantial runaway velocity,
suggest that a recent binary merger is the best explanation for the 
rapid rotation of VFTS~102.  An example of a possible merger progenitor 
is the nearby contact binary VFTS~352 with an orbital period of 1.1 d
\citep{Almeida2015}.

\subsection{VFTS 285}  

VFTS~285 is among some ten objects that appear to be fleeing from the 
R136 cluster at the center of 30~Dor \citep{Evans2010,Lennon2018, Platais2018,Renzo2019,Gebrehiwot2021}.  
These are all examples of ejection by SN or dynamical encounters. 
VFTS~285 has a relative tangential velocity in the range of 26 to 48
km~s$^{-1}$ with a time of flight since ejection of 0.6 to 0.7 Myr if it
originated in the R136 complex at the center of the NGC 2070 cluster
\citep{Platais2018,Gebrehiwot2021}.
We find that the radial velocity is $250 \pm 6$ km~s$^{-1}$ that is
only somewhat smaller than the mean for its cluster of origin, NGC~2070,
$271 \pm 12$ km~s$^{-1}$ \citep{Evans2015}, so both the radial and 
tangential velocities are consistent with the idea that VFTS~285 is 
a ``slow'' runaway star. 

There are several factors to consider in determining how VFTS~285 was
ejected. \citet{Schneider2018} estimate an age of $1.9 \pm 2$ Myr for
VFTS~285 (see \citealt{Platais2018}) that is consistent with its
location in the HRD near the ZAMS position of a track for its mass (Fig.~12).  If this is the actual age, then it is too young for sufficient
time to have elapsed for the companion to evolve and explode as a SN
(at least 3 Myr for the most massive stars).  This young age would 
indicate instead that the star was ejected by dynamical processes in the
R136 complex that has a similar age ($\approx 1.2$ Myr;
\citealt{Bestenlehner2020}). On the other hand, VFTS~285 may have been
ejected from another site in the NGC~2070 cluster, which has a median
age of 3.6~Myr \citep{Schneider2018}. In this case, there is sufficient
time for a binary companion to reach the SN stage and eject VFTS~285,
and then its estimated age would correspond to that of the rejuvenated
star after mass accretion in the binary.  

The other fact to note is the very high He abundance we determined for 
VFTS~285 ($y=0.34 \pm 0.14$).  This level of He enrichment is not
predicted by single star evolution for a star of its mass and youth, but
it could happen through mass transfer from an evolved companion or by
large scale mixing associated with a merger.  The large He abundance
implies an interaction with some kind of evolved object, so the
binary path is probably the most likely one for the evolutionary history
VFTS~285. It may have been spun up through mass transfer from a
companion that exploded and disrupted the binary (like the case of $\zeta$~Oph).  
\citet{Bestenlehner2020} found that the most massive WN5h-type stars 
in the core of R136 are all He enriched at their surfaces even
at very young ages ($\approx 1.2$ Myr), so mass transfer from such 
a progenitor companion could potentially explain the He overabundance in VFTS~285. 
There is one other system that 
may be a post-mass transfer binary in the 30 Dor region. 
\citet{Clark2015} found that the rapid rotator VFTS~399 is 
a strong emitter of variable X-ray emission that is usually 
associated with thermal emission from an accreting neutron star, 
so they suspect that VFTS~399 is a high-mass X-ray binary. 
This would indicate that there has been enough time in some parts
of 30 Dor for a possible binary companion of VFTS~285 to explode as 
a SN, so the formation channel by a SN disruption of the original 
binary is a viable and attractive explanation. 

We doubt that VFTS~285 is the result of a merger, because it would
have formed from two lower mass, longer-lived stars, which conflicts with the young age derived from its kinematical and cluster properties.  It is possible, however, 
that a dynamical encounter with another binary led to the ejection of a 
binary with such high eccentricity that the pair collided and merged on an 
orbital timescale. Thus, the merger scenario should not be entirely
ruled out. 


\section{Conclusions} 

VFTS~102 and VFTS~285 are the current record
holders for the fastest projected equatorial
velocity, and our findings have only solidified
that standing.  We applied a spectrum synthesis
method to create model spectral line profiles
that depend on the physical parameters and the 
inclination angle $i$ between the spin axis 
and our line-of-sight.  
Fits of these to the observed profiles 
led to determinations of both the projected 
rotational velocity $v_{e} \sin i$ and the physical
equatorial velocity $v_e$ (Table~4). 
We found that both stars are exceptionally fast rotators 
with VFTS~102 rotating at $v_e = 649$ km~s$^{-1}$
($\Omega / \Omega_{c} = 1.00$) 
and VFTS~285 rotating at $v_e = 648$ km~s$^{-1}$ 
($\Omega / \Omega_{c} = 0.95$). 
The physical parameters 
associated with our best fit for VFTS~102 are: 
$R_{p}/R_\odot = 5.41 \pm 1.55$, $M/M_\odot = 18 \pm
6$, and $T_{p} = 40100 \pm 2800$ K. 
VFTS~285 is slightly larger and more massive: 
$R_{p}/R_\odot = 5.58 \pm 0.39$, $M/M_\odot = 28 \pm 8$, 
and $T_{p} = 40200 \pm 2700$ K. 

Both stars exhibit very broad and shallow (and 
often blended) line profiles due to the extreme 
rotational line broadening.  We calculated 
models for 16 line or line blend features, 
and from these we selected three H Balmer, 
three \ion{He}{1}, and three \ion{He}{2} lines 
that could be modeled with a self-consistent 
set of parameters.  We found that the best fit
parameters also led to predicted profiles for 
UV features that matched well with the observed, 
strongly blended spectra.  The temperature related 
variations of surface specific intensities have a 
much greater contrast between the pole and equator 
in the UV than in the optical, so 
the profile shapes are expected to differ 
\citep{Hutchings1976}.  Thus, it is encouraging 
that our models which incorporate the wavelength 
dependence of specific intensity are generally 
successful in fits of both the UV and optical lines. 

As a part of our analysis we attempted to measure
the He abundance after it became clear that the
models associated with a solar He abundance were
far too weak to match our observed spectra.
We found that all of the target stars are 
He overabundant ($\zeta$~Oph,
$2.4 \times$ solar; VFTS~102, $2.0
\times$ solar; VFTS~285, $3.4 \times$
solar). This general He overabundance is 
probably the result of internal mixing 
promoted by extreme rotation and/or by  
past mass transfer of He from an 
evolved mass donor companion. 
We caution that while these stars all appear He enriched,
the actual He abundances may have systematic errors because
the fits were made by simply increasing the He abundance in 
the radiative transfer solution for the line profiles 
without re-calculating the full atmospheric structure 
for the revised He abundance (see Section 4.3).

A comparison of the stellar parameters to those
of evolutionary tracks for rapid rotators \citep{Brott2011}
shows that both VFTS~102 and VFTS~285 appear 
to be somewhat overluminous for their mass.  
Furthermore, both stars are much more enriched 
in He than predicted by mixing in these model tracks.
These characteristics as well as their implied 
youth suggest that both VFTS~102 and VFTS~285 
may have been rejuvenated by mass transfer from 
an interacting binary companion.  Their current fast 
rotation may be the result of angular momentum 
accretion during past mass transfer. 

VFTS~102 is rotating very close to the critical rate, 
is shedding mass and angular momentum into a 
circumstellar disk, and is enriched in He. 
These are all the characteristics of a recent, 
post-merger object as suggested by \citet{Jiang2013}.
The star's radial velocity and proper motion 
are similar to those of 
the nearby OB-association LH~99, 
so we doubt that VFTS~102 is a runaway star 
(as suggested by \citealt{Dufton2011}). 

VFTS~285, on the other hand, does appear to be 
a runaway star ejected from the R136 cluster 
based upon its proper motion \citep{Platais2018}.
An attractive scenario is that VFTS~285 was spun up 
via mass transfer prior to its companion exploding in a
supernova. The binary was disrupted and the orbital 
motion of the survivor was transformed into the 
linear ejection velocity of VFTS~285. 
In this picture, the extreme overabundance of He 
in the atmosphere of the star marks the remains 
of nuclear-processed gas from the interior of 
the mass donor companion.  
This scenario is similar to the current
origin theory for $\zeta$ Oph \citep{Neuhauser2020}, 
which has a similar He abundance to that of VFTS~285.
Our work adds to a growing body of evidence that 
a significant fraction of the rapid rotators among the 
massive stars were spun up through binary mass transfer
\citep{Bodensteiner2020,Wang2021,Gies2022}.

\begin{acknowledgments}
We are grateful to Nolan Walborn (deceased) and Denise Taylor of STScI 
for their aid in planning the observations with HST. Support for program GO-14246 was provided by NASA through a grant from the Space Telescope Science Institute, which is operated by the Association of Universities for Research in Astronomy, Incorporated, under NASA contract NAS5-26555. Some of the data presented in this paper were obtained from the Mikulski Archive for Space Telescopes (MAST). Support for MAST for non-HST data is provided by the NASA Office of Space Science via grant NNX13AC07G and by other grants and contracts. This work has made use of data from the European Space Agency (ESA) mission
{\it Gaia} (\url{https://www.cosmos.esa.int/gaia}), processed by the {\it Gaia} Data Processing and Analysis Consortium (DPAC,
\url{https://www.cosmos.esa.int/web/gaia/dpac/consortium}). Funding for the DPAC has been provided by national institutions, in particular the institutions participating in the {\it Gaia} Multilateral Agreement.
Additional support was provided from the National Science Foundation under grant AST-1908026 and from the GSU College of Arts and Sciences.
\end{acknowledgments}

\vspace{5mm}
\facilities{CFHT (ESPaDOnS), HST (COS), IUE, VLT:Kueyen (X-shooter, FLAMES)}

\software{TLUSTY/SYNSPEC}


\bibliography{ms1}{}
\bibliographystyle{aasjournal}

\end{document}